\documentclass[aps,prd,amsmath,floats,floatfix, twocolumn,
nosuperscriptaddress,nofootinbib,letterpaper]{revtex4}

\usepackage[colorlinks, pdfborder={0 0 0}, plainpages=false]{hyperref}
\usepackage{graphicx}
\usepackage{xspace}
\usepackage[usenames,dvipsnames]{color}
\usepackage{amssymb}
\usepackage{microtype}
\usepackage{bm}

\newcommand{\Oberlin}{\affiliation{Department of Physics and Astronomy,
    Oberlin College, Oberlin, Ohio 44074, USA}} %

\newcommand{\dA}{\hspace{1mm} dA}
\newcommand{\SpEC}{{\tt SpEC}}

\begin{document}

\title{
Black Hole Spin Axis in Numerical Relativity
}

\author{Robert~Owen}\Oberlin
\author{Alex~S.~Fox}\Oberlin
\author{John~A.~Freiberg}\Oberlin
\author{Terrence~Pierre~Jacques}\Oberlin

\hypersetup{pdfauthor={Owen et al.}}

\date{\today}

\begin{abstract}
Colliding black holes are systems of profound interest in both gravitational wave astronomy and in gravitation theory, and a variety of methods have been developed for modeling their dynamics in detail. The features of these dynamics are determined by the masses of the holes and by the magnitudes and axes of their spins. While masses and spin magnitudes can be defined in reasonably unambiguous ways, the spin axis is a concept which despite great physical importance is seriously undermined by the coordinate freedom of general relativity. Despite a great wealth of detailed numerical simulations of generic spinning black hole collisions, very little attention has gone into defining or justifying the definitions of the spin axis used in the numerical relativity literature. In this paper, we summarize and contrast the various spin direction measures available in the {\tt SpEC} code, including a comparison with a method common in other codes, we explain why these measures have shown qualitatively different nutation features than one would expect from post-Newtonian theory, and we derive and implement new measures that give much better agreement.
\end{abstract}

\maketitle

\section{Introduction}

The age of gravitational-wave astronomy has arrived. In the short time since it began operation, the Laser Interferometer Gravitational-Wave Observatory (LIGO) has detected three clear gravitational-wave events from binary black hole mergers~\cite{GW150914, GW151226, GW170104}, with many more expected to follow, and a revolutionary multi-messenger observation of colliding neutron stars~\cite{GW170817}. After decades of effort, researchers are probing the strong-field dynamics of spacetime itself, with steadily increasing reach and precision. 

Like all cutting-edge science, gravitational wave physics is a continuous conversation between observational and theoretical efforts. 
One theoretical technique of particular importance for exploring the strong-field dynamics of spacetime is {\em numerical relativity}, the direct computational solution of Einstein's field equations~\cite{alcubierreBook, baumgarteShapiroBook}. Numerical relativity can provide a picture of spacetime dynamics with no approximations other than the usual truncation error of numerical calculation, which in principle is straightforward to control. 

Unfortunately, along with the exact treatment of spacetime geometry that numerical relativity provides, there also arises a great deal of ambiguity associated with the ``general covariance'' at the heart of Einstein's theory. For mathematical analysis, all fields are represented in some coordinate system, however (with some technical caveats) the theory is fundamentally ambivalent about what coordinate system is used. Standard approximation techniques such as the post-Newtonian expansion~\cite{Blanchet2006, Futamase:2007zz} and black hole perturbation theory~\cite{Sasaki:2003xr} assume the existence of a subset of preferred coordinate systems in which deviations of the spacetime metric from its stationary state are ``small.'' In numerical relativity, one often makes vague statements about whether a coordinate system (a ``gauge'', in this context) is ``good'' or ``bad'', but generally very little effort goes into formalizing such statements beyond what is necessary to expect a stable evolution and sensible behavior of the evolving fields.

One of the quantities computed in numerical relativity of particular physical importance is {\em black hole spin}. The parameter space of non-eccentric binary black hole systems is seven-dimensional --- described by the ratio of the holes' masses, and the three spin components on each hole. Other parameters, such as the total mass of the system, the time of the merger, and the distance to the detector, are important for data analysis but fundamentally unimportant for source modelling as they can be altered trivially in post processing. The detections that LIGO has made thus far have claimed precise measurements of the black hole masses, but relatively rough measurements of the black hole spins and their states of alignment. Black hole spin is a phenomenon that imprints itself less strongly on gravitational waves than black hole mass, but it is a key near-term target for precision measurement as LIGO's sensitivity improves. 

A rich and increasingly relevant body of literature exists on treating binary black hole systems with arbitrarily-aligned spins in numerical relativity, exploring basic dynamical processes such as spin flips~\cite{Campanelli2007a, Campanelli2007b}, remnant ``kicks'' due to asymmetric wave generation~\cite{Baker2006c, Gonzalez2007, Bruegmann-Gonzalez-Hannam-etal:2007, Tichy:2007hk, Baker2008, Healy2008, Lousto:2011kp}, and of course the increasingly crucial work of numerical relativity groups at filling out ``catalogs'' of binary black hole waveforms~\cite{Mroue:2013PRL, SXSCatalog, GaTechCatalog, RITCatalog} and tuning approximate waveform models to numerical results~\cite{Field2014, Purrer2015, Blackman2017a, Blackman2017prl, Blackman2017b}. Beyond this, a plethora of dynamical effects associated with spin alignment have been studied, a non-comprehensive list of which would include~\cite{Lousto:2013wta, Lousto:2013, Lousto:2014, Zlochower:2015, LoustoHealyNakano, LoustoHealy2016}. 

While the spin axis is clearly a crucial element of modern numerical relativity simulations, with an increasingly strong connection to gravitational wave data analysis, its basic {\em definition} is somewhat vague and generally ad-hoc in numerical relativity treatments. Much of this centers on its inherent gauge ambiguity: it is defined by its alignment with the coordinate axes of the simulation, assuming an unphysical Euclidean background geometry. While the spin {\em magnitude} can be defined and practically computed in a number of gauge-invariant ways~\cite{OwenThesis, Cook2007, Lovelace2008, Lovelace:2014}, the spin axis requires greater care and finesse. In the long run, one can hope that the physical content that is currently described with the ``spin axis'' might be replaced with more basic, gauge-invariant concepts such as relationships among horizon multipoles~\cite{Ashtekar2004, Owen2009, AshtekarCampigliaShah}, for the near-term it is important to at least be specific about what is meant by ``spin axis'' in various codebases, and to be aware of the peculiarities of specific measures of the axis. 

In particular, the definition of the spin axis used in the \SpEC~code~\cite{SpECwebsite} is nontrivial and has never been described in the literature, though multiple papers have been written which used this measure at a fundamental level. In particular, in Ref.~\cite{Ossokine2015} the precession dynamics in \SpEC~simulations were compared with expectations from post-Newtonian theory (PN). While most features agreed quite well between numerical relativity and post-Newtonian theory, the spins of the individual black holes were found to nutate in a surprising way, qualitatively different from the expectations of PN theory. In this paper, we will argue that this unexpected nutation behavior can be traced back to a peculiarity of {\SpEC}'s default spin-axis measure (and, to our knowledge, the measures used in all recent numerical relativity calculations), and can be removed with the use of a different measure of the spin axis.

The paper is organized as follows: in Sec.~\ref{s:background}, we lay out some mathematical preliminaries that will be useful in the further discussion. In Sec.~\ref{s:existingtechniques}, we summarize existing definitions of the black hole spin axis, including the definition employed in the \SpEC~code. Along the way, we summarize a few of the theoretical motivations (boost invariance, centroid invariance) that led to its introduction. Then, in Sec.~\ref{s:discrepancy}, we compare these various spin measures in the case of a nontrivially-precessing binary black hole merger, and  we describe the surprising nutation features by which the \SpEC~spin measure differs qualitatively from PN results. In Sec.~\ref{s:newmeasure}, we employ a more straightforward technique, defining the spin axis by a quasilocal angular momentum formula using {\em coordinate rotation generators}. This straightforward method has significant theoretical drawbacks, which we will outline (relegating some particularly tedious details to an appendix), but we find that the practical ambiguities are minimal, and some of them (boost ambiguity) can be understood and mitigated through more careful analysis of the underlying mathematics. This provides a new measure of the black hole spin axis which preserves some of the advantageous features of~\SpEC's previous measure, while agreeing much better with post-Newtonian expectations. Finally, in Sec.~\ref{s:discussion}, we summarize these results and outline some prospects for future work.

\section{Mathematical preliminaries}
\label{s:background}

The standard techniques for computing black hole spin in the modern 
numerical relativity literature, including the techniques of this paper, 
are founded upon the following {\em quasilocal angular momentum} formula:
\begin{equation}
J = \frac{1}{8 \pi_0} \oint \vec \omega \cdot \vec \phi \dA, \label{e:SpinFormula}
\end{equation}
where the integral is carried out over a spatial two-surface normally 
of spherical topology (in practice, an apparent horizon), $\vec \phi$ is 
some rotation-generating vector field, and $\vec \omega$ is the 
normal-tangential 
projection of the (undensitized) canonical momentum conjugate to the spatial 
metric:
\begin{equation}
\omega_\mu = (K_{\rho \nu} - K g_{\rho \nu}) h_\mu^\rho s^\nu,
\end{equation}
where $\vec s$ is the unit normal to the integration 2-surface, tangent to the  
spatial slice, $h_\mu^\rho = \delta_\mu^\rho + u_\mu u^\rho - s_\mu s^\rho$ is the projector 
tangent to this 2-surface ($u^\mu$ is the timelike normal to the spatial slice), $K_{\mu \nu}$ is the extrinsic curvature of the 
spatial slice, 
and $g_{\mu \nu} = \psi_{\mu \nu} + u_\mu u_\nu$ is the {\em spatial} metric (the spacetime metric $\psi$ projected 
down to the spatial slice). The clunky introduction of a subscript in $\pi_0 := 3.14159...$ is due to the fact that it will eventually become convenient to use the letter $\pi$ to denote a potential associated with momentum.

There are various justifications for this quasilocal angular momentum formula 
in the literature. It arises in the quasilocal charge constructions of Brown 
and York~\cite{BrownYork1993}. It naturally arises again in the formalisms of 
isolated and dynamical horizons~\cite{Ashtekar1999, Ashtekar2001, Ashtekar-Krishnan:2004}, though there the 
``spatial slice'' is often taken to be the horizon worldtube (which is spacelike for a dynamical horizon). The 
result is the same, though, for either slicing, so long as the rotation 
generator is chosen in such a way as to make $J$ ``boost invariant'', as we 
will discuss below. This angular momentum formula can also be 
shown to be eqivalent to the Komar angular momentum in axisymmetric spacetimes~\cite{Komar:1959}. 

Because both $\vec \omega$ and $\vec \phi$ are defined to be tangent to the 
2-surface, it is natural to write Eq.~\eqref{e:SpinFormula} in a basis that 
makes this explicit:
\begin{equation}
J = \frac{1}{8 \pi_0} \oint \omega_A \phi^A \dA,\label{e:SpinFormulaBasis}
\end{equation}
where capital latin letters index the 2-surface tangent bundle. In this paper we will frequently consider the manner in which quantities transform with respect to the ``boost gauge'' freedom. That is: the freedom to alter the slicing of spacetime {\em near} the 2-surface, while leaving the 2-surface itself fixed. For this, it is convenient to introduce a Newman-Penrose null tetrad~\cite{Newman1962}, $l^\mu, n^\mu, m^\mu, \bar m^\mu$, where $l^\mu n_\mu = -1$, $m^\mu \bar m_\mu = 1$, and all other dot products vanish. Furthermore, we will adapt this tetrad to the 2-surface, as in the Geroch-Held-Penrose (GHP) variant of the formalism~\cite{Geroch1973, Penrose1992}, such that the real and imaginary parts of $\vec m$ are tangent to the 2-surface, and the real null vectors $\vec l$ and $\vec n$ are respectively its outgoing and ingoing null normals. Finally, we fix the remaining scaling freedom in $\vec l$ and $\vec n$ by adapting them to the timelike normal to the slicing, $\vec u$, and the spacelike normal to the 2-surface, within the slicing, $\vec s$, such that:
\begin{align}
\vec l &= \frac{1}{\sqrt{2}} \left( \vec u + \vec s\right), & \vec n &= \frac{1}{\sqrt{2}} \left( \vec u - \vec s\right),\\
\vec u &= \frac{1}{\sqrt{2}} \left( \vec l + \vec n\right), & \vec s &= \frac{1}{\sqrt{2}} \left( \vec l - \vec n\right).\label{e:usfromln}
\end{align}
Substituting these formulas into the standard formula for the extrinsic curvature of the slicing:
\begin{equation}
K_{\mu \nu} = - g_\mu^\rho g_\nu^\sigma \nabla_\rho u_\sigma,
\end{equation}
where $g_\mu^\rho = \delta_\mu^\rho + u_\mu u^\rho$ is the projector to the spatial slice, it is straightforward to express the normal-tangential projection, $\vec \omega$, in terms of the tetrad legs. Specifically, $\omega_A$ represents a connection on the bundle of normal vectors to the 2-surface in spacetime:
\begin{equation}
\omega_A = e_A^\rho n_\sigma \nabla_\rho l^\sigma.\label{e:ConnectionOnNormalBundle}
\end{equation} 
While this form will be most convenient for our purposes, it is worth noting 
that this expression can be massaged further into standard GHP spin 
coefficients:
\begin{equation}
\omega_A = \left(\beta^\prime - \bar \beta \right) m_A + \left( \bar {\beta^\prime} - \beta \right) \bar m_A.
\end{equation}
The combinations of coefficients shown here are indeed the ``connection 
coefficients'' that relate the tetrad derivative operator 
$\delta := m^\mu \nabla_\mu$ to the GHP derivative operator $\eth$ for 
quantities $Q$ of spin-weight zero and boost-weight $\pm 1$, namely, 
components of spacetime vectors normal to the 2-surface:
\begin{equation}
\eth Q = \delta Q \pm \left(\bar {\beta^\prime} - \beta\right) Q.
\end{equation}

Because $\omega_A$ geometrically represents a connection, it is natural (and 
will be of practical use below) to define its corresponding curvature. 
\begin{align}
\Omega &= \epsilon^{AB} \nabla_A \omega_B\label{e:OmegaDefined}\\
&= -i m^{[A} \bar m^{B]} \nabla_A \omega_B
\end{align}
The scalar quantity $\Omega$ is the imaginary part of the ``complex curvature'' 
of a 2-surface embedding, defined by Penrose and Rindler~\cite{Penrose1992}:
\begin{equation}
\Omega = \Im \left[ \sigma \sigma^\prime - \rho \rho^\prime - \Psi_2 + \Phi_{11} + \Lambda\right],
\end{equation}
where $\rho$ and $\rho^\prime$ are the GHP coefficients representing the 
complex expansions of $\vec l$ and $\vec n$, respectively, $\sigma$ and 
$\sigma^\prime$ are the coefficients representing the shears, $\Phi_{11}$ and 
$\Lambda$ are components of the spacetime Ricci curvature, which we will take 
to vanish in this paper (due to Einstein's field equations and the assumption 
of vacuum), and $\Psi_2$ is the ``Weyl scalar'' that is intuitively taken 
to represent the 
non-radiative part of the spacetime Weyl tensor. (Though rigorous statements 
upon those lines require the choice of a special tetrad. See, for example, 
~\cite{Beetle2005, Zhang:2012ky}.) The real part of the complex 
curvature, which we will not 
invoke here, is the familiar gaussian curvature of the 2-surface. It should be 
noted that $\sigma^\prime$ and $\rho^\prime$ vanish on an isolated horizon, and 
thus in vacuum $\Omega$ is simply the imaginary part of $\Psi_2$, up to a sign.
Furthermore, the quantity $\Im\left[\Psi_2\right]$ in this basis is precisely 
the normal-normal component of the magnetic part of the spacetime Weyl tensor,
$B_{ss} = B_{ij} s^i s^j$, which is referred to intuitively as the 
``horizon vorticity'' in Refs.~\cite{OwenEtAl:2011, Nichols:2011pu, Zimmerman2011, ZhangPRD2, VortexPRD3}.

The ``boost gauge'' transformation that often appears in discussion of quasilocal quantities is a transformation that leaves the 2-surface unchanged but boosts the spatial slice around it. If the timelike normal to the spatial slice, $\vec u$, and the spacelike normal to the 2-surface within the spatial slice, $\vec s$, are defined as in Eqs.~\eqref{e:usfromln}, then this boost is easily described by a rescaling of the null normals:
\begin{align}
\vec l &\mapsto \exp(a) \hspace{1mm} \vec l,\\
\vec n &\mapsto \exp(-a) \hspace{1mm} \vec n,
\end{align}
where $a$ is some scalar on the 2-surface, representing the rapidity of the boost at each point.

Under such a transformation, it is readily seen from Eq.~\eqref{e:ConnectionOnNormalBundle} that the quasilocal angular momentum density $\vec \omega$ is {\em not} invariant, but rather transforms as:
\begin{equation}
\omega_A \mapsto \omega_A - \nabla_A a
\end{equation}
However, because this correction term is a pure gradient, the curvature quantity $\Omega$, defined in Eq.~\eqref{e:OmegaDefined}, is boost invariant.

The boost invariance of $B_{ss}$ is even simpler to argue, since it is simply the imaginary part of the Weyl scalar $\Psi_2$. In terms of tetrad vectors and the spacetime Weyl tensor $C_{\mu \nu \rho \sigma}$, $\Psi_2$ is:
\begin{equation}
\Psi_2 = \frac{1}{2} C_{\mu \nu \rho \sigma} \left(l^\mu n^\nu l^\rho n^\sigma - l^\mu n^\nu m^\rho \bar m^\sigma\right),
\end{equation}
which is manifestly invariant under boost transformations.

\section{Existing techniques for defining the spin axis}
\label{s:existingtechniques}

A basic assumption underlying essentially all methods for computing black hole 
spin in binary systems is that the holes can in some sense, at least 
approximately, be considered isolated from their partners and from the 
dynamics of the surrounding spacetime. This intuitive idea invites 
appeals to the 
structure of the Kerr geometry, which would be expected to accurately 
represent a nearly isolated, uncharged, vacuum black hole. The event 
horizon of the Kerr geometry can be foliated 
by marginally outer trapped surfaces, which 
for a Kerr black hole would coincide with the 2-surfaces found by the 
``apparent horizon finder.'' Furthermore, as long as the slicing of spacetime 
conforms to the global axisymmetry of the Kerr geometry, the apparent horizon 
2-surface would then also be expected to be axisymmetric, with the axisymmetry 
describing a symmetry under rotations about the spin axis. Hence, under the 
assumption that the Kerr geometry is an accurate approximation of the spacetime 
near the horizon (or, quasilocally speaking, that the horizon itself is 
``isolated'' in the sense of~\cite{Ashtekar1999, Ashtekar-Krishnan:2004}), 
then the axisymmetry of the horizon can be used to define the spin 
axis.\footnote{Note that throughout the remainder of this paper we 
will adopt the common parlance of numerical relativity and use the word 
{\em horizon} to refer to the two-dimensional marginally trapped surface 
computed by the code's horizon finder.} 

\subsection{Euclidean line between poles or extrema}
\label{s:Extrema}
To our knowledge, all binary black hole codes 
other than~\SpEC~infer the spin axis from an axis of best approximate horizon symmetry 
(and an option along these lines is available in~\SpEC~as 
well, as we will outline below, though it is not the default measure). 
Specifically, the most common technique 
applies methods outlined in~\cite{Dreyer2003}, in which a Killing vector field 
is found on a black hole horizon (or some kind of approximation of a Killing 
field if an exact one does not exist) by integrating the Killing transport 
equations, a system of ODEs that must be satisfied by a Killing vector 
field along any given path.

More precisely: the method first identifies a {\em candidate} Killing vector at a {\em point} on the horizon. To do this, a three-dimensional vector space of data is constructed at the starting point. Then the Killing transport equations are used to propagate each of the basis vectors around a {\em closed} path. In so doing, the starting vectors are mapped to new vectors in the same tangent space in which they started. This mapping is linear, so one can compute corresponding eigenvalues and eigenvectors. The eigenvector with eigenvalue closest to unity is considered to be the best candidate for a Killing vector field over the whole horizon, because a true Killing vector would indeed return to itself under this transport. Once such a ``best'' vector is found at the chosen starting point, it is propagated to the rest of the horizon grid again using Killing transport. All published results that we're aware of that identify an approximate Killing vector (AKV) field do so using either this method or the method described below that's used in the~\SpEC~code. In particular, this technique is the default behavior of the {\tt QuasilocalMeasures} thorn of the {\em Einstein Toolkit}~\cite{EinsteinToolkit}.

Once such a rotational Killing vector field has been constructed on the 
horizon, its poles (the isolated points where the computed 
vector field has zero norm, of which there are hopefully two) can be used to distill a kind of spin axis vector 
as:
\begin{equation}
\hat \chi_{KT}^i := (x_2^i - x_1^i)/N, \label{e:PoleConnector}
\end{equation}
where $N$ is a normalization factor chosen to make 
$\delta_{ij}\hat \chi_{KT}^i \hat \chi_{KT}^j = 1$,
and $x_1^i$ and $x_2^i$ (for $i=1,2,3$) are the global Cartesian coordinate 
values of the two poles. Note that this does not in itself define a ``spin 
vector'', but rather a unit-norm ``axis vector'' (unit-norm in the Euclidean 
background space). We will use the same $\hat \chi$ notation for other axis 
vector definitions described later in this paper. In order to define a 
``spin vector'', be it an angular momentum vector or some variant rescaled by 
some power of the mass, one must multiply this axis vector by an appropriate 
magnitude. The obvious (and ubiquitous) choice is the spin magnitude defined 
by Eq.~\eqref{e:SpinFormula}, where $\vec \phi$ is taken to be the approximate 
Killing vector field. 

Note also that beginning with Eq.~\eqref{e:PoleConnector} we are 
making essential reference to the ``background'' coordinates of the numerical 
simulation. This is an ambiguous procedure. In principle, one could  
remap the spatial coordinates to achieve an arbitrary change in the 
$\hat \chi_{KT}$ vector. This will unfortunately be a pattern in all of the 
spin axis prescriptions described in this paper. The spin axis vector as 
conventionally understood in numerical relativity is not a true geometric 
object, but rather defined explicitly in terms of a particular Euclidean 
background geometry. Like the stress-energy pseudotensor of Landau and
Lifshitz~\cite{Landau-Lifshitz}, it transforms covariantly under 
global Poincar\'e transformations of the spatial coordinates, however a 
nonlinear coordinate transformation would change the underlying background 
geometry and thereby change the meaning of $\hat \chi_{KT}$. In 
the numerical relativity literature, there is a general hope that the 
simulation coordinates --- normally chosen for code stability and/or 
computational convenience --- are nonetheless ``well adapted'' to the dynamics 
of the horizon. In practice, this naive hope is often rewarded. 
For example, after a black hole merger, one would expect all fields to settle 
to stationary values --- that is, for the coordinate time-translation vector 
to settle to the stationarity Killing vector of the eventual Kerr geometry. 
Indeed this occurs, to our knowledge, under all of the standard gauge 
conditions of modern numerical relativity codes. Moreover, the spatial 
coordinates not only adapt to the late-time stationarity, but at least in 
the~\SpEC~code the time slicing even settles to conventional slicings of 
black-hole perturbation theory, with horizon multipoles ringing at 
precisely the frequencies calculated in perturbation theory~\cite{Owen2009}. 
Furthermore, after a black hole merger, the spacetime describing the 
remnant black hole is axisymmetric in simulation coordinates, even though there 
is no obvious reason why the coordinate gauge conditions should ``find'' the 
geometrical symmetries of the simulated spacetime. 
This ``good behavior'' of the simulation coordinates is still a somewhat 
mysterious phenomenon, deserving of deeper analysis. This mystery should 
encourage caution in the use of any background-dependent methods. But 
nonetheless, these are the techniques that are ubiquitous in the field, and 
with some practical justification.

In principle, it may be possible to describe the essential elements of the 
black hole spin axis in truly unambiguous language, for example using the 
formalism for ``source multipoles'' on dynamical horizons defined 
in Ref.~\cite{AshtekarCampigliaShah}. Here, data on the horizon is projected 
against test functions that evolve on the dynamical horizon in such a way as 
to represent a ``fixed'' frame of reference, in a specific sense. We intend to 
explore this approach in future work, but for now we simply note that its 
restriction to data on the horizons themselves would cloud efforts at 
connecting the evolution of the two distinct holes, or of describing the 
relationship of spin precession to dynamics of the encompassing spacetime, as 
one often wishes to do in binary black hole simulations. Thus, the techniques 
used throughout the current paper stick with background-dependent methods 
as described above. 

The axis vector $\hat \chi_{KT}$ is not easily adaptable to the~\SpEC~code, 
because the approximate Killing vector field constructed by the Killing 
transport method does not lead to a smooth vector field in the limit of an 
infinitely-refined grid. 
Many of the calculations in~\SpEC~require smooth fields, due to the 
pseudospectral numerical methods that suffuse the code. 

A different 
definition of approximate Killing vector fields, used ubiquitously in 
the~\SpEC~code, was presented in~\cite{OwenThesis, Lovelace2008} (essentially 
the same technique was independently presented in~\cite{Cook2007}). Briefly 
summarizing: the approximate Killing field $\vec \xi$ is first presumed to 
have zero divergence on the horizon, as required by the trace of Killing's 
equation:
\begin{equation}
D_A \xi^A = 0,
\end{equation}
where $D_A$ is the covariant derivative on the horizon 2-surface. Such a 
vector field can easily be constructed from a scalar potential $\zeta$:
\begin{equation}
\xi^A = \epsilon^{AB} D_B \zeta,
\end{equation}
where $\epsilon^{AB}$ is the 2-surface Levi-Civita tensor. Finally, a condition 
is imposed on $\zeta$ that it minimize the integrated squared shear of 
$\vec \xi$ 
over the 2-surface --- that is, the remaining residual of Killing's equation. 
This condition leads to a simple generalized eigenproblem for $\zeta$ on the 
horizon. This eigenproblem reduces to the eigenproblem of the horizon Laplacian 
on a metrically-round 2-sphere, and hence $\zeta$ can be considered a kind of 
generalized spherical harmonic, an idea explored further in~\cite{Owen2009}. 
If the horizon is not strongly deformed, the eigenfunction with the lowest 
corresponding eigenvalue (that is, the one corresponding to an AKV with the 
smallest integrated squared shear) will also align with the spin axis, when 
we can roughly define this axis by independent means. Hence, the vector 
$\xi^A = \epsilon^{AB} D_B \zeta$ serves the same purpose as the vector defined 
by the Killing transport method above, though it is a smooth vector field and 
an ``approximate'' solution to Killing's equation in a specific variational 
sense. 

When the approximate Killing vector field $\vec \xi$ is defined in this way, 
its poles are the extrema of the scalar function $\zeta$. This fact provides 
a natural connection of the machinery in the~\SpEC~code to the $\hat \chi_{KT}$ 
technique. While the approximate Killing vector field computed in~\SpEC~is not 
the same vector field computed in the Killing transport method, we can 
nonetheless carry out an analogous procedure to define and study a very similar 
axis measure $\hat \chi_{\zeta e}$, defined by:
\begin{equation}
\hat \chi_{\zeta e}^i = (x_2^i - x_1^i)/N,\label{e:NonSpECDefault}
\end{equation}
where again $N$ is simply a normalization factor, but $x_2^i$ and $x_1^i$ are 
the coordinates of the extrema of $\zeta$. 

This procedure, while mathematically straightforward, becomes slightly tricky 
in~\SpEC, where all horizon data is resolved into a pseudospectral 
expansion in spherical harmonics. The high accuracy of the spectral expansion 
allows the code to run with rather coarse grids. In particular, in the 
runs presented in this paper, the spectral expansion of horizon data is 
resolved up to spherical harmonic order $\ell = 15$. This implies a typical 
spacing between collocation points of 
$\Delta \phi = 2 \pi_0/(2 \ell + 2) \approx .2$ rad $\approx 11^\circ$. In this 
paper, we intend to probe nutations of the spin axis to much higher precision 
than that. Thus, in finding extrema of $\zeta$ (or of other functions, as we'll 
discuss below), we must either employ an analytic formula to find locations of 
extrema purely from spectral coefficients (and we know of no such formula), or 
we must carry out a search procedure that probes data interpolated between the 
grid points of the horizon. To this end, our code first finds extrema of 
$\zeta$ restricted to the coarse grid of collocation points. It then carries 
out a straightforward gradient-descent algorithm on the interpolated values of 
$\zeta$ to locate extrema between collocation points. We have found this 
technique to be robust only when the extrema are not near the poles of the 
horizon's spherical coordinate chart, however the simulations presented here 
satisfy this requirement. Because, as we will see, the use of extrema provides 
no particular advantage over the techniques already employed in~\SpEC~(and 
provides an inferior measure of spin axis to the methods used later 
in this paper), we have not attempted to improve this technique to make it more 
robust for general calculations. 

This method of defining a spin axis by the Euclidean line between extrema of 
a function can also be extended to other relevant quantities on the horizon. 
In particular, the quantities $\Omega$ and $B_{ss}$ are more directly related 
to quasilocal angular momentum and frame-dragging than $\zeta$ is, and 
therefore might be expected to be less susceptible to tidal effects. In this 
paper, we will also explore quantities called $\hat \chi_{\Omega e}$ and 
$\hat \chi_{B_{ss} e}$, both defined similarly to $\hat \chi_{\zeta e}$, but 
with $x_1^i$ and $x_2^i$ referring to minima and maxima of $\Omega$ and $B_{ss}$.

\subsection{Angular momentum vector from coordinate rotation generators}
\label{s:SimpleCoordSpin}
Early in the development of the modern~\SpEC~code, we explored methods along 
the lines described above, but were concerned that the holes, which raise tides 
on one another, might deform enough that their best symmetry axis 
might not be determined solely by the spin axis, but also by the orientation 
of the tidal bulge. For this reason, we explored a more basic approach, 
defining the spin axis through integrals of the form of 
Eq.~\eqref{e:SpinFormula}. Specifically, it was our hope that a spin ``vector'' 
could be defined as:
\begin{equation}
J_i = \frac{1}{8 \pi_0} \oint_{\cal H} \vec \omega \cdot \vec \phi_{i} \dA,\label{e:CoordSpin}
\end{equation}
where ${\cal H}$ refers to the horizon 2-surface, and the $\vec \phi_{i}$ are 
coordinate rotation vectors. The most familiar definition of the coordinate rotation vectors is the following:
\begin{equation}
\vec \phi_{i} = \eta_{ij}{}^k \left(x^j - x_0^j\right) \vec \partial_k.\label{e:CoordinateRotationGenerators1}
\end{equation}
Here $\eta_{ijk}$ represents the alternating tensor of the flat ``background'' 
geometry, a totally antisymmetric object with $\eta_{123} = 1$ in the 
cartesian, asymptotically-inertial coordinate system of the simulation, and $\vec \partial_k$ represent the coordinate translation vectors of this coordinate system. Throughout this paper, indices $i,j,k,...$ refer specifically to the basis associated with the background coordinate system, and are raised and lowered trivially with the flat metric $\delta_{ij}$. Indices $a,b,c,...$ refer to the basis associated with some arbitrary coordinate system, and are raised and lowered with the physical spatial metric $g_{ab}$ like standard coordinate indices. The three constant quantities $x_0^j$ represent {\em centroid} coordinates associated with the particular horizon under consideration.

It is worth noting that another reasonable definition of the coordinate rotation generators can be used, constructed from translation {\em one-forms}:
\begin{equation}
\bm{\varphi}^{i} = \eta^i{}_{jk} \left(x^j - x_0^j\right) {\bf d}x^k.\label{e:CoordinateRotationGenerators2}
\end{equation}
Though these one-forms are very closely related to the vectors in Eq.~\eqref{e:CoordinateRotationGenerators1}, they are {\em not} the same if the physical spatial metric does not coincide with the background metric. That is to say:
\begin{equation}
\phi_{i}^a \neq \delta_{ij} g^{ab} \varphi^{j}_b
\end{equation}
In Appendix~\ref{a:TheTwoGenerators}, we will explore the relationship between 
these two coordinate rotation generators. 

Either rotation vector is dependent on the choice of centroid $x_0^k$.
There are many ways to fix this centroid, a fact that we'll come back to in 
detail in Sec.~\ref{s:newmeasure}, and different conditions for $x_0^k$ would 
be expected to lead to 
different angular momentum integrals. The most obvious one, with which we 
experimented the most in the early days of the~\SpEC~code, used horizon 
averages of the coordinates:
\begin{equation}
x_0^i = \frac{1}{A} \oint_{\cal H} x^i \dA.
\end{equation}

Unfortunately, the spin measure defined in this way showed some features that 
do not comport with the behaviors normally associated with black 
hole spin. The most glaring of these was a feature in which a supposedly 
``nonspinning'' black hole (in an equal-mass binary where neither hole 
initially has spin) settles (after the initial burst of junk radiation) 
to a hole with spin in the direction opposite the orbital angular momentum. 
This much is perfectly plausible as a real, physical effect --- junk radiation 
can be absorbed by the holes 
and can cause them to spin up slightly. The difficulty comes in the 
later inspiral, in which this small (yet numerically well resolved) initial 
spin, 
anti-aligned with the orbital angular momentum, {\em increases} during the 
ensuing inspiral. One could expect tidal viscosity effects to spin up initially 
nonspinning holes, but such spinup would be {\em aligned} with the orbital 
angular momentum in a system like this one. By this measure, 
an initially nonspinning hole appears to spin up 
during inspiral in the direction {\em opposite} the expectations from 
perturbative calculations. This phenomenon can be seen in 
Fig.~\ref{f:WrongWaySpinup}, in which it can also be noted that the effect is 
strongly influenced by the choice of centroid $x_0^i$.

\subsection{Axis vector defined by coordinate moments}

Given the strange behaviors of $J_i$ described in Sec.~\ref{s:SimpleCoordSpin}, and the uncertainties of choosing the centroid 
$x_0^i$, a decision was made early in the development of~\SpEC~to deemphasize 
this spin vector and to instead define the spin axis in a manner somewhat 
analogous to the methods of Sec.~\ref{s:Extrema}. 
However, to avoid 
the subtleties of rootfinding described there, we instead defined the spin 
axis through coordinate moment integrals of spin-related quantities on the 
horizon. In analogy with $\hat \chi_{\zeta e}$, 
one could define a similar spin axis through coordinate moments as:
\begin{equation}
\hat \chi^i_{\zeta m} := \frac{1}{N}\oint_{\cal H} \zeta x^i \dA.
\end{equation}
Or, again, under the assumption that $\Omega$ or $B_{ss}$ might respond  
differently to tidal effects, one could define spin axes through their 
moments:
\begin{align}
\hat \chi^i_{\Omega m} &:= \frac{1}{N}\oint_{\cal H} \Omega x^i \dA,\label{e:SpECDefaultMeasure}\\
\hat \chi^i_{B_{ss} m} &:= \frac{1}{N}\oint_{\cal H} B_{ss} x^i \dA.
\end{align}
These spin axis vectors are all defined as unit vectors, not as full angular 
momentum vectors, and thus are undefined in the case of zero spin. It  
therefore wouldn't make sense to ask if they share the strange 
nonphysical spinup features of the $J_i$ in Sec.~\ref{s:SimpleCoordSpin}. 
However, the non-normalized forms of $\hat \chi^i_{\Omega m}$, 
and $\hat \chi^i_{B_{ss} m}$ do indeed vanish (to within the estimated 
truncation error of the simulation) for the same equal-mass nonspinning 
runs.\footnote{$\hat \chi^i_{\zeta m}$ does not have a ``non-normalized'' form, 
because $\zeta$, defined by an eigenvalue problem, has no definite scale 
unless some normalization condition is applied to it.} The three measures 
described here have two other critical features, which also happen to be 
shared with the extremum measures defined in Sec.~\ref{s:Extrema}:
\begin{itemize}
\item \textbf{\em Centroid invariance:} If the spatial coordinates are offset 
by constants, $x^i \mapsto x^i + \Delta x^i$, the spin axes $\hat \chi^i_{\zeta m}$,
$\hat \chi^i_{\Omega m}$, and $\hat \chi^i_{B_{ss} m}$ do not change. This is 
because the quantities $\zeta$, $\Omega$, and $B_{ss}$ each have zero mean when 
integrated over the 2-surface, and thus if the $\Delta x^i$ are constants, the 
additional terms integrate to zero.
\item \textbf{\em Boost-gauge invariance:} As described in 
Sec.~\ref{s:background}, it is useful for theoretical reasons that the spin 
measure be unchanged under slicing transformations that leave the 2-surface 
itself fixed. For one, this invariance ensures that arguments made about the 
spin evaluated in the code slicing also apply to the spin as evaluated on 
the spacelike {\em dynamical horizon} 3-surface. All of the quantities that 
appear in the above moment integrals are boost-gauge invariant. $\zeta$ is 
defined intrinsically to the 2-surface, so it is manifestly boost-gauge 
invariant. The invariance of $\Omega$ and $B_{ss}$ was argued in Sec.~\ref{s:background}. 
\end{itemize}

The measure $\hat \chi_{\Omega m}$ is the default measure of the spin axis used 
in the~\SpEC~code, and to our knowledge has been the sole measure of spin 
direction in all publications of~\SpEC~results. More precisely,~\SpEC~outputs 
a ``spin vector'', which is simply the ``unit vector'' $\hat \chi_{\Omega m}$, 
multiplied by the dimensionless spin magnitude computed by approximate Killing 
vector methods~\cite{Lovelace2008}.

It should be noted that boost-gauge invariance does not mean the spin 
direction measure is {\em slicing} invariant. Indeed, one would not expect an 
angular momentum vector to be slicing invariant~\cite{Schiff1960}. If we 
consider a Kerr black hole, represented in, say, Kerr-Schild coordinates
$x^\mu = (t, x, y, z)$, and then reslice the spacetime using a global Lorentz 
boost, then the shape of the horizon, represented in the boosted spatial 
coordinates on the boosted time slice, will be length contracted along the 
direction of the boost by a factor $1/\gamma$. Since the spin axis vectors 
defined above (at least before normalization) are linear in the spatial 
coordinates $x^i$, the components of these axis vectors along the direction 
of the boost are reduced by the same factor. This transformation means that the 
{\em angle} that the angular momentum vector makes with the direction of motion 
varies with spin and boost speed in precisely the same way as in special 
relativity and post-Newtonian theory under the Pirani spin-supplementary 
condition~\cite{Schiff1960}. We have confirmed this transformation behavior 
by evaluating these spin axis measures on boosted spinning Kerr black holes 
in~\SpEC.

Also, while the technique outlined here to define $\hat \chi_{\Omega m}$ through 
coordinate moments may seem ad-hoc, it actually arises from 
Eq.~\eqref{e:SpinFormula} in a straightforward manner. On a sphere of radius 
$r_0$ in Cartesian coordinates in a Euclidean geometry, the rotation vectors 
can be rewritten as:
\begin{equation}
\varphi^{i}_A = r_0 \epsilon_A{}^B D_B x^i,
\end{equation}
where the three coordinates $x^i$ are treated as scalars with regard to the 
covariant derivative $D$ on the 2-surface. If we use this vector field to 
evaluate the quasilocal angular momentum using Eq.~\eqref{e:SpinFormulaBasis}, 
then an integration by parts gives:
\begin{align}
J^{i} &= \frac{r_0}{8 \pi_0} \oint \omega_A \epsilon^{AB} D_B x^i \dA,\\
&= \frac{r_0}{8 \pi_0} \oint x^i \epsilon^{BA} D_B \omega_A \dA,\\
&= \frac{r_0}{8 \pi_0} \oint x^i \Omega \dA.
\end{align}
In practice, of course, the horizon is {\em not} a round sphere embedded in 
Euclidean space, so this is simply a motivating argument, not any kind of 
derivation. But this argument, along with the centroid and boost-weight 
invariance of $\hat \chi_{\Omega m}$, were the main motivating factors that 
led to its adoption as the default measure of spin axis in~\SpEC.

\section{Discrepancies between these measures, and with Post-Newtonian results}
\label{s:discrepancy}

The previous section outlined six different options for defining the 
spin axis in numerical relativity, each about equally well motivated. All 
involve scalar quantities on the horizon (specifically, the scalar curvature, 
$\Omega$ of the horizon's normal bundle, which very naturally distills the 
boost-invariant information from the quasilocal angular momentum density; 
the normal-normal component of the magnetic part of the Weyl tensor, $B_{ss}$, 
called the ``horizon vorticity'' in Ref.~\cite{OwenEtAl:2011}, which equals 
$\Omega$ when the horizon is stationary, and can be easily related to 
differential frame dragging; and the potential $\zeta$ associated with an 
approximate symmetry of the horizon). Further, these scalar quantities could be 
distilled into an ``axis'' either by taking integral moments over the horizon 
with respect to the background Cartesian coordinates, or by tracing a 
coordinate line between their extrema. The~\SpEC~code, by default, uses 
moments of $\Omega$, while most non-\SpEC~ papers trace a 
line between the poles of the approximate symmetry vector (a technique that 
we will model here with lines between the extrema of $\zeta$). We have also mentioned, and will explore in greater detail below, the fact that one could simply insert coordinate rotation generators into the angular momentum formula, a strategy that unfortunately presents even more subtleties and ambiguities.

Given the many options for defining the spin axis, it would be helpful to 
calculate the discrepancies among these measures in a representative case, 
as a rough measure of how well they 
might be trusted as measures of the specific physical quantity that they are 
all meant to represent. This section is devoted to such a comparison.

\subsection{Estimate of numerical uncertainty}

Because the spin axis measures we describe below will in some cases differ by very small amounts, we must be careful to account for the inherent numerical uncertainty of the calculations. The~\SpEC~code uses a rather elaborate system of adaptive mesh refinement, so detailed convergence analysis is not straightforward. The situation for the calculations in this paper is even more complicated, as many of our calculations involve calculus on the horizon itself, which is itself resolved to some finite spectral order, again chosen adaptively. 

As a rough measure of numerical truncation error, one can simply measure the angle between corresponding axis measures in the two highest-resolution calculations described in this paper. Figure~\ref{f:NaiveConvergence} shows such a comparison for the particular measures $\hat \chi_{\Omega m}$ (coordinate moments of $\Omega$), $\hat \chi_{\zeta m}$ (moments of $\zeta$), $\hat \chi_{\Omega e}$ (extrema of $\Omega$), $\hat \chi_{\zeta e}$ (extrema of $\zeta$), and the normalized $\hat J$ computed using the coordinate rotation vectors of Eq.~\eqref{e:CoordinateRotationGenerators1}. This image implies that one could expect the spin axis measures quoted in this paper to be accurate to within a fraction of a degree over most of the inspiral. It should come as no surprise, given the numerical issues involved, that the axis measures defined by extrema are particularly noisy, especially at very early times as the horizons ring in response to the absorption of junk radiation. A close analysis of these extremum-based curves also show occasional ``glitches'' at which the spin axis jumps discontinuously. The locations of extrema can jump discontinuously even for a function that changes smoothly, so discontinuous jumps could conceivably be physical in origin, however we don't claim to have isolated them in any numerically well-defined sense. Thankfully, over most of the inspiral these unresolved discontinuities are smaller in magnitude than the error from direct comparison of continuous segments (of order $0.1^\circ$ for $\hat \chi_{\Omega e}$, and slightly smaller for $\hat \chi_{\zeta e}$. The bottom three curves in Fig.~\ref{f:NaiveConvergence}, describing measures based on integrals over the horizon, show significantly less error early in the simulation.

\begin{figure}
 \includegraphics[width=\columnwidth]{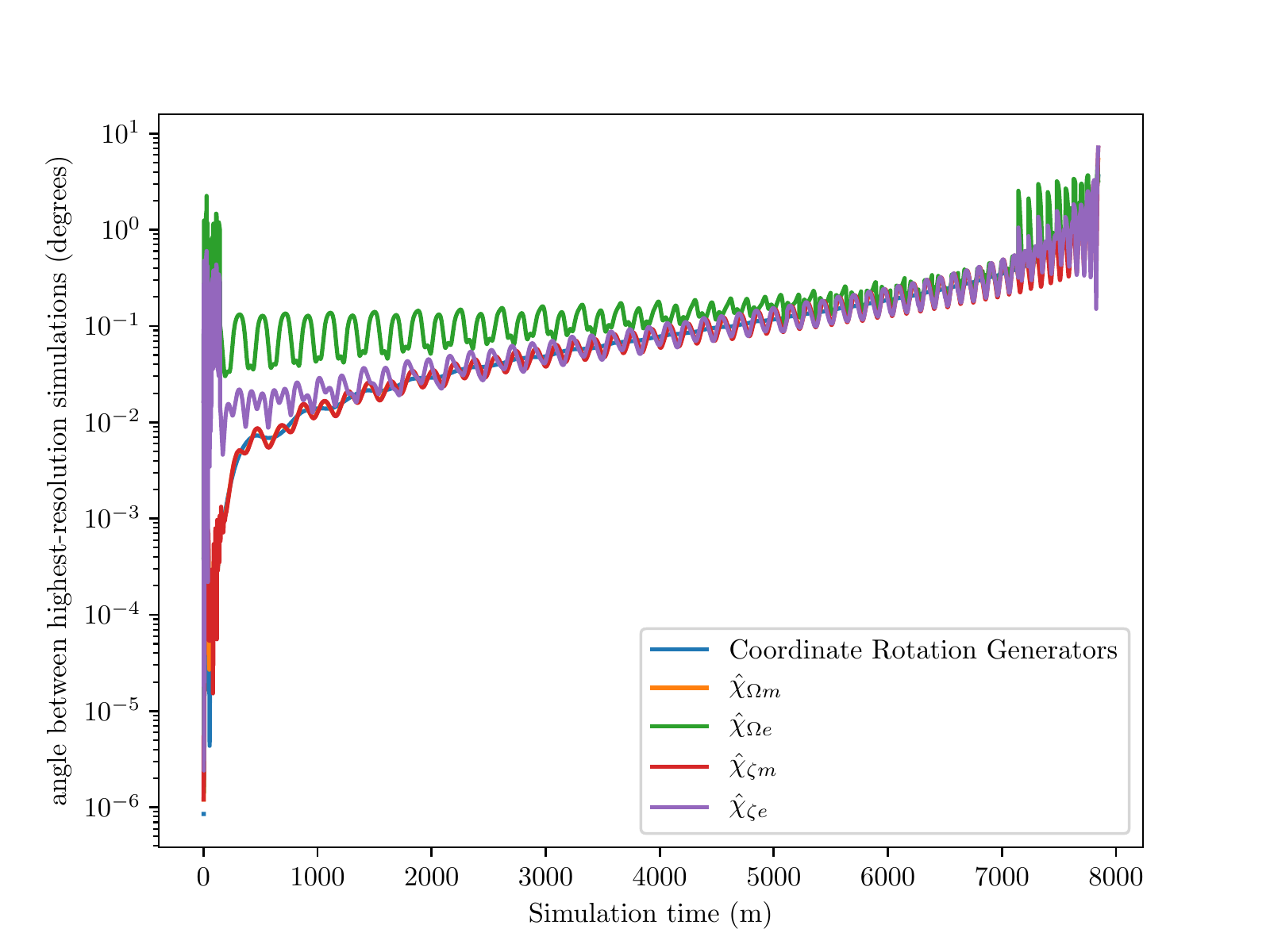}
 \caption{\label{f:NaiveConvergence} Angle, in degrees, between spin axis measures computed in two highest resolution simulations of the merger described in this section, for five different such measures. The top two curves show estimated error in measures based on the axis between extrema of $\Omega$ or $\zeta$. The lower curves show errors in spin axis measures based on the coordinate rotation vectors $\vec \phi_{i}$ defined in Eq.~\eqref{e:CoordinateRotationGenerators1} (blue, smoother), and those inferred by coordinate moments of $\Omega$ and $\zeta$ (orange and red, respectively, and mostly overlapping on this scale).
}\label{f:NaiveConvergence}
\end{figure}

Another feature in the bottom four curves of Fig.~\ref{f:NaiveConvergence} is worth noting: they appear to show roughly the same accumulation of error over the course of the inspiral despite their being computed in very different ways. One might naturally conclude that the accumulated error in these three spin axis measures is not primarily due to truncation error in the spin axis calculation, but rather due to accumulated error in the {\em orbital phase} over the course of the inspiral. Such accumulated phase error would create an {\em apparent} dephasing of the spin, simply because the comparisons between different-resolution simulations in Fig.~\ref{f:NaiveConvergence} are made at matching {\em coordinate times} rather than at matching orbital phase. Such dephasing would cause the higher-resolution spin axis to systematically lead or lag that at lower resolution, even if the overall precession tracks agree much better. 

\begin{figure}
 \includegraphics[width=\columnwidth]{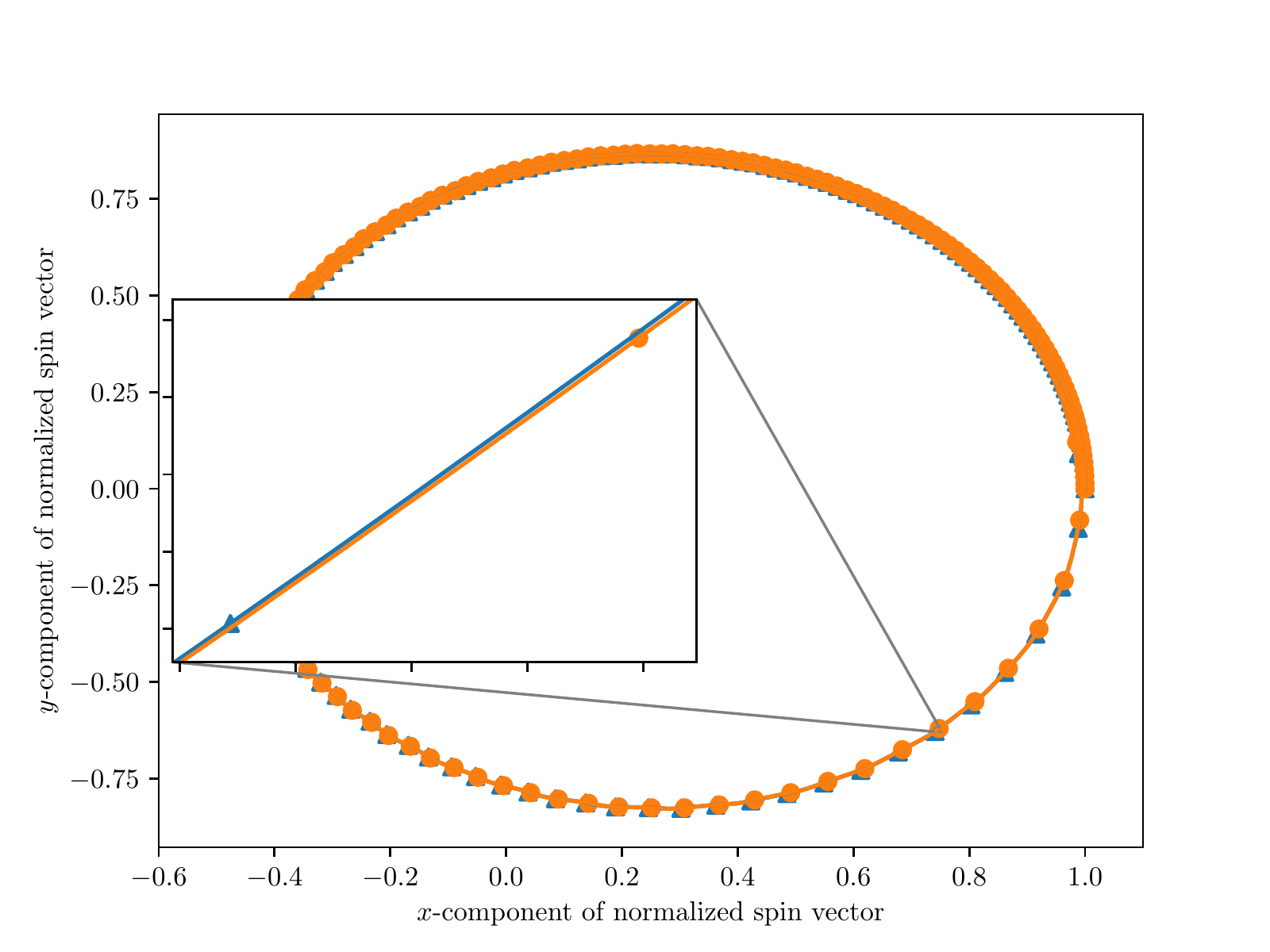}
 \caption{\label{f:XYTrack} Parametric plot of $x$ and $y$ components of normalized spin axis vector computed from coordinate rotation generators defined in Eq.~\eqref{e:CoordinateRotationGenerators1}, for our two highest-resolution simulations, with a blowup showing that the tracks themselves follow one another more closely than the individual data points at equal times. Analogous graphs are qualitatively similar for all of our spin axis measures. 
}
\end{figure}

\begin{figure}
 \includegraphics[width=\columnwidth]{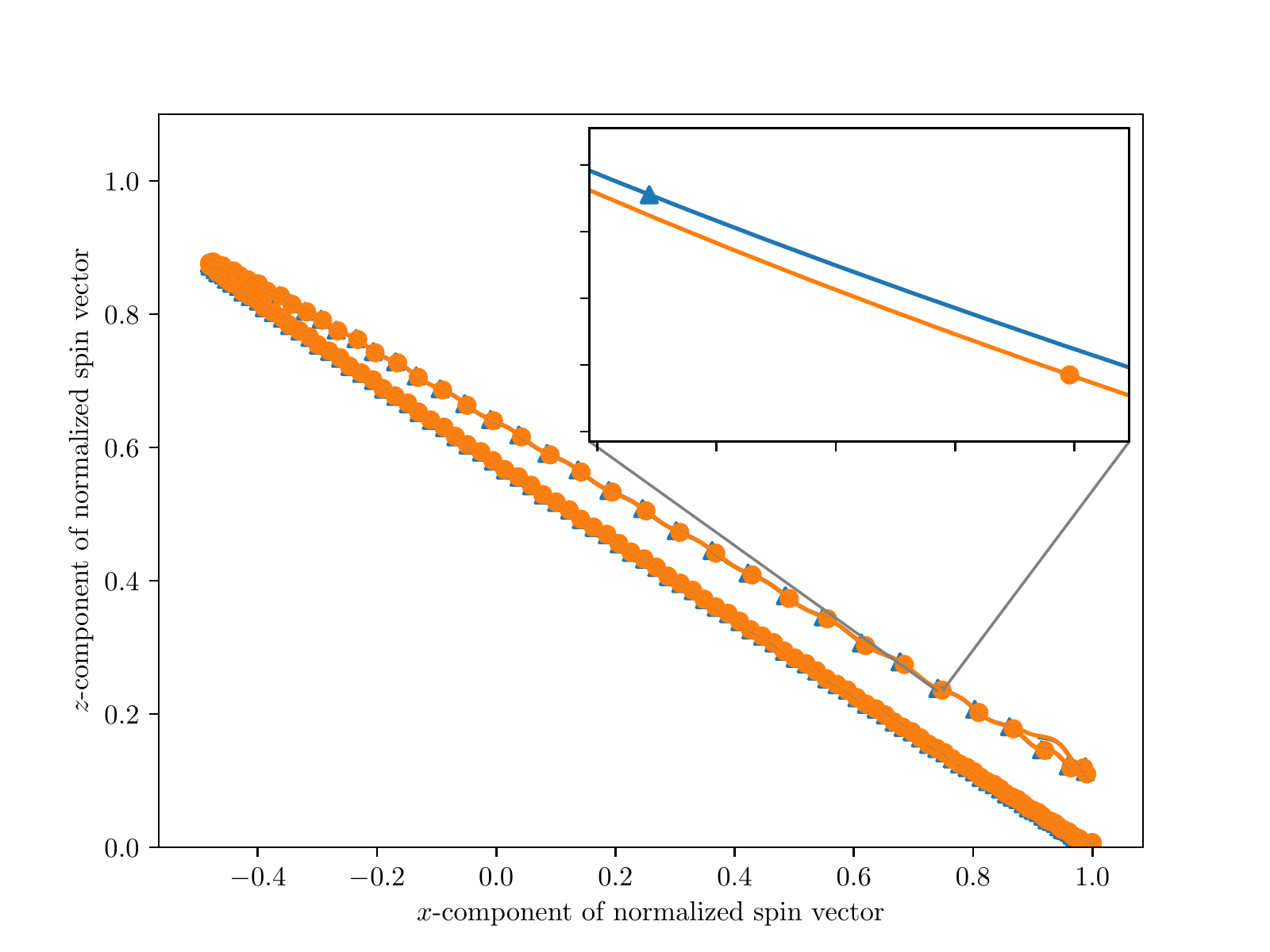}
 \caption{\label{f:XZTrack} Parametric plot of $x$ and $z$ components of normalized spin axis vector computed from coordinate rotation generators defined in Eq.~\eqref{e:CoordinateRotationGenerators1}, for our two highest-resolution simulations, with a blowup showing that the tracks themselves follow one another more closely than the individual data points at equal times. Analogous graphs are qualitatively similar for all of our spin axis measures. 
}
\end{figure}

In Figs.~\ref{f:XYTrack}--\ref{f:XZTrack}, we zoom in on a region of the spin precession track, on a slice through the 3-d vector space of spin vectors (scaled to unit norm in the global vector space, to ease the inference of angular discrepancies). Markers on the lines show data for specific measurement times. As expected, the distance between the two tracks is significantly less than the distance of the labeled grid points, indicating that timing delay seems to dominate the error shown in Fig.~\eqref{f:NaiveConvergence}, at least at late times. A more optimistic estimate of the axis uncertainty would be the typical shortest angular distance between spin curves at the two highest resolutions, which we estimate to be roughly $.02$ degrees for measures based on function moments ($\hat \chi_{\omega m}$, $\hat \chi_{B_{ss} m}$, $\hat \chi_{\zeta m}$) or any of the coordinate rotation generators. This estimate is also consistent with the angular distances shown at early times in Fig.~\ref{f:NaiveConvergence}. 

An optimistic estimate of the axis error is shown in Fig.~\ref{f:OptimisticAngleConvergence}. Here, as a rough means of correcting for possible orbital dephasing, at each timestep of the high-resolution simulation the spin axis is compared to {\em all} timesteps of the second-highest-resolution simulation, and the smallest relative angle, in degrees, is shown in Fig.~\ref{f:OptimisticAngleConvergence}. We emphasize that this measure is both {\em rough} (there is discretization error from the fact that both simulations have data dumped only every 0.5$M$ of coordinate time), and possibly {\em optimistic} (the comparison made here would be insensitive to relative precessions if they happen to follow the same precession track). Nonetheless, relative angles below $.01^\circ$ throughout most of the run (until, we presume, faster precession makes the discretization error more problematic) add further weight to our rough estimate of $.02^\circ$ errors in all spin measurements except those based on function extrema (for which we stick to a more conservative estimate of $0.1^\circ$). 

\begin{figure}
 \includegraphics[width=\columnwidth]{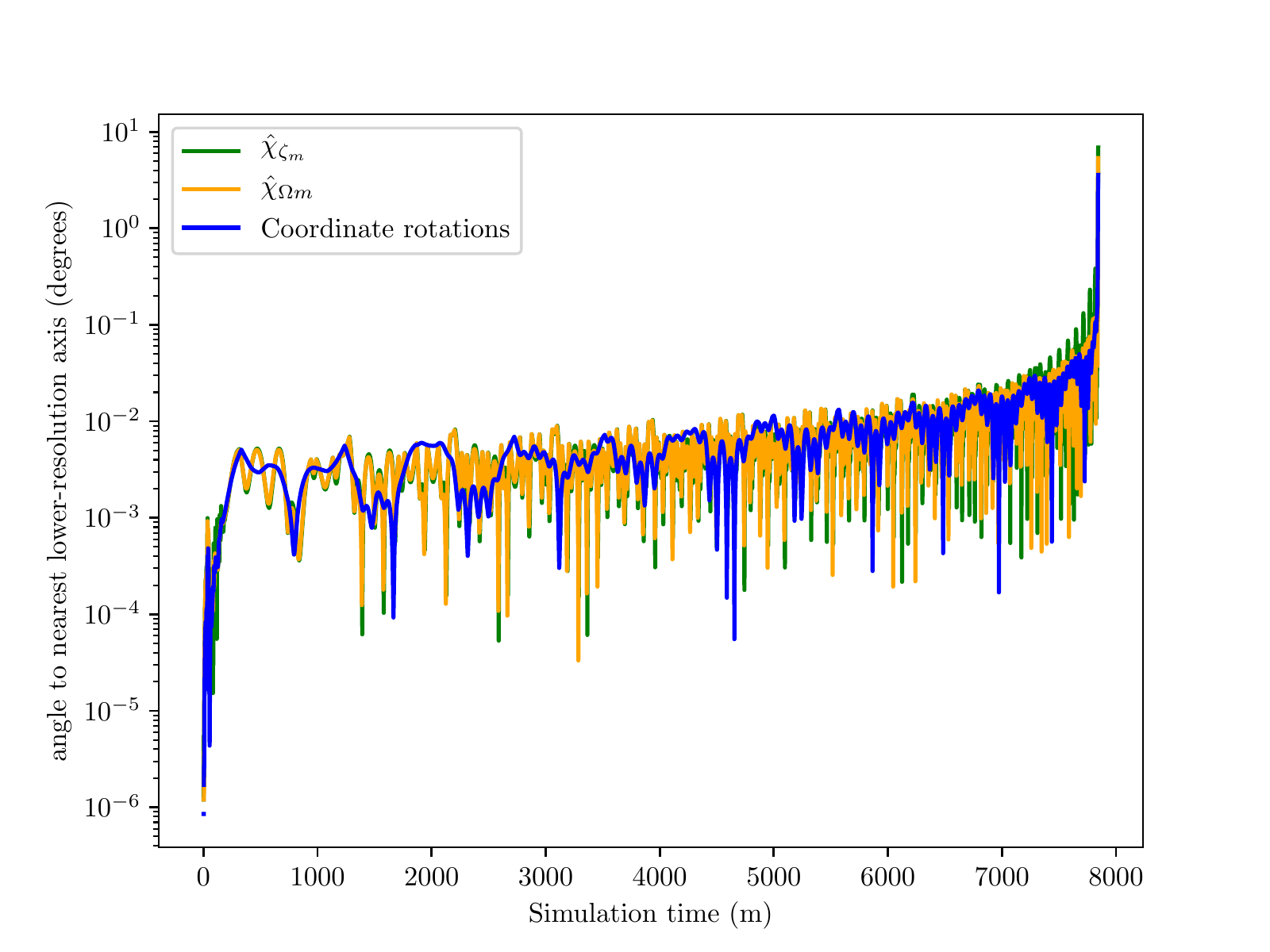}
 \caption{\label{f:OptimisticAngleConvergence} An optimistic representation of the numerical error in the integral-based axis measures. Each timestep of the highest resolution run is compared to the best of all {\em time-delayed} counterparts in the second-highest resolution. 
}\label{f:OptimisticAngleConvergence}
\end{figure}

\subsection{Analysis of a particular case}
We now consider a merger which involves simple though nontrivial 
nutation properties. This simulation is of particular interest because the 
nutations were found (using the standard 
measure from the~\SpEC~code, $\hat \chi_{\Omega m}$) to display features 
that are qualitatively inconsistent with expectations from Post-Newtonian 
theory~\cite{Ossokine2015}. The case is a low-eccentricity inspiral 
of black holes with mass ratio 5:1, in which the less massive hole is 
nonspinning and the more massive hole has spin magnitude 0.5$m_1^2$ (where $m_1$ is 
the mass of this larger hole, 5/6 the total mass $m$ of the system), with 
initial 
spin direction (according to the $\hat \chi_{\Omega m}$ measure) tangent to the 
initial orbital plane. This simulation was referred to as {\tt q5\_0.5x} in 
Ref.~\cite{Ossokine2015}, and further details can be found 
at~\cite{SXSCatalog} where this configuration is numbered 
0058, though for this study we reran the simulation to 
add the new spin measures.

\begin{figure}
 \includegraphics[width=\columnwidth]{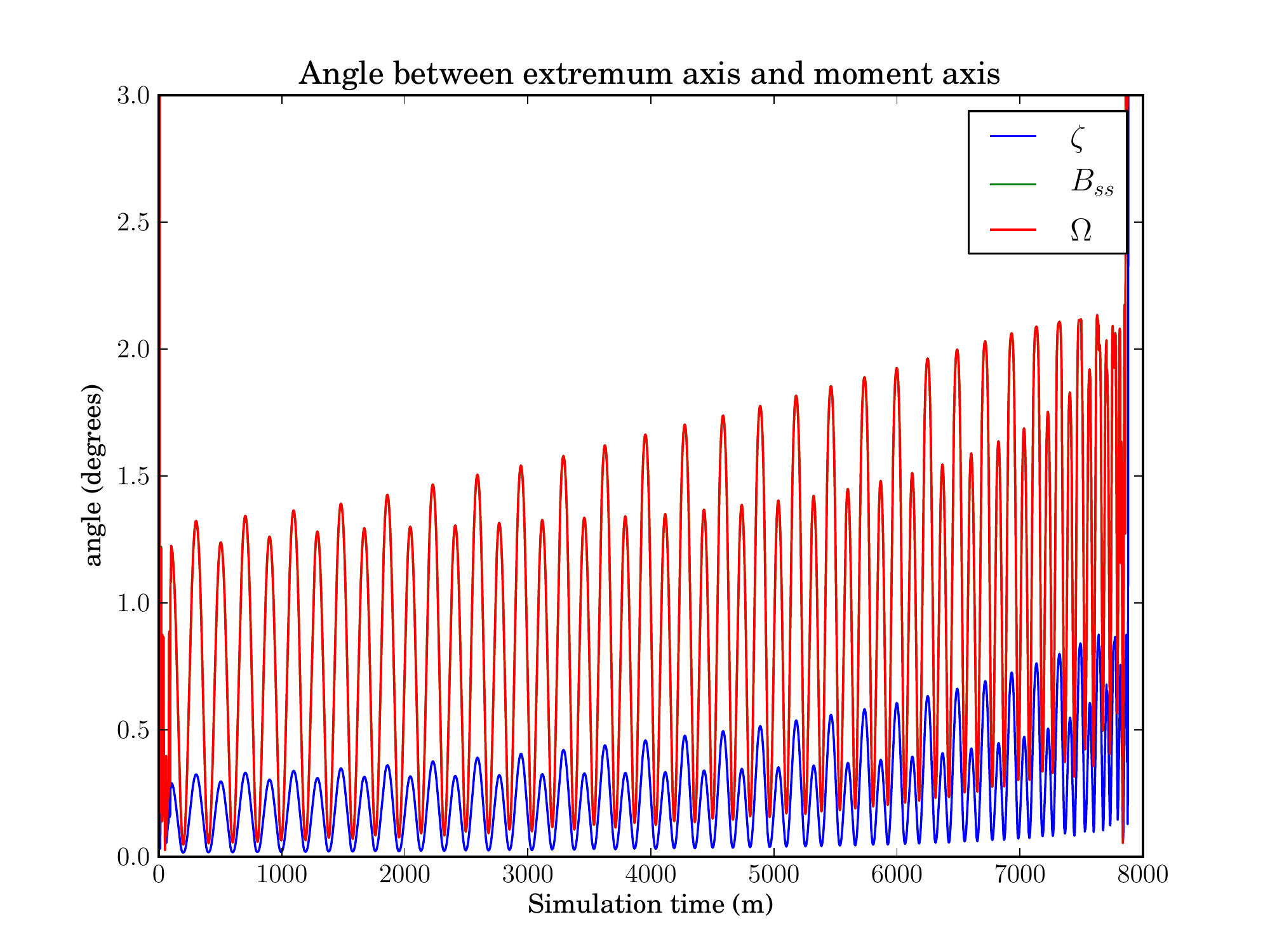}
 \caption{\label{f:ExtremaVsMoments} Angle (in the Euclidean background space) 
between spin axes defined using coordinate moments and using line between 
extrema. The data for the horizon vorticity $B_{ss}$ (green curve) and the 
closely related curvature $\Omega$ of the normal bundle (red curve), overlap 
so precisely that only one can be seen in this figure. For all three scalars, 
agreement is to within about a degree throughout the inspiral. 
For reference, the orbital period in the early inspiral is 
approximately 400$m$, so the oscillations in these curves are on the timescale 
that one would expect from the rotation of a tidal bulge.
}
\end{figure}

\begin{figure}
 \includegraphics[width=\columnwidth]{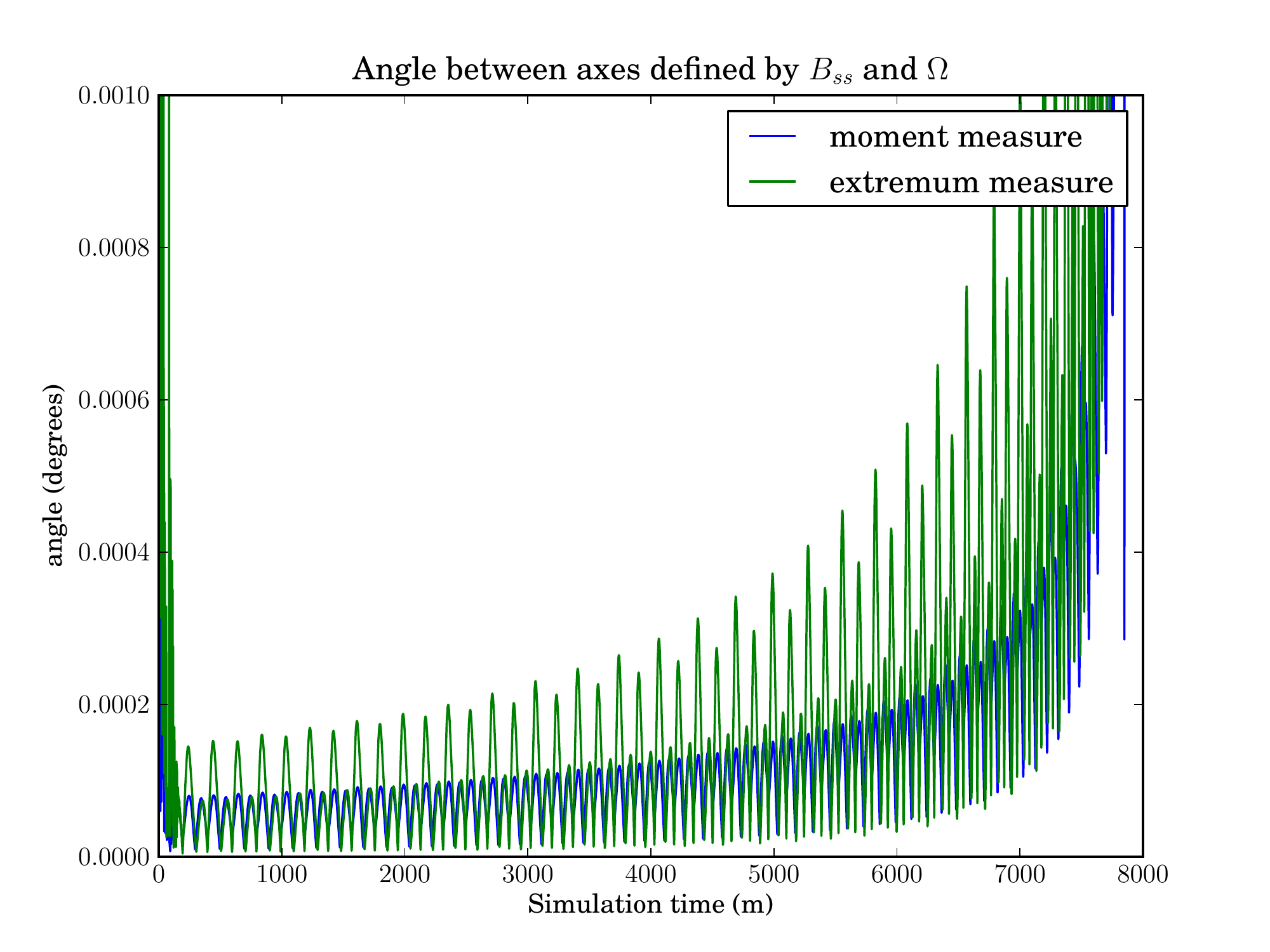}
 \caption{\label{f:BssVsOmega} Angle (in the Euclidean background space) 
between spin axes defined using $B_{ss}$ and $\Omega$, either using coordinate 
moments or a line between extrema. The measures agree remarkably well, a sign 
that $B_{ss}$ is very nearly equal to $\Omega$, as one would expect for holes 
that are not undergoing strong dynamical processes. In fact, the measures agree 
to within a hundredth of a degree right up to the formation of the common 
horizon. It should be noted that we estimate the numerical truncation error 
of both of these measures to be roughly on the order of $0.02^\circ$, 
significantly greater than either of these discrepancies. Thus, even 
these differences could be largely due to numerical truncation.}
\end{figure}

\begin{figure}
 \includegraphics[width=\columnwidth]{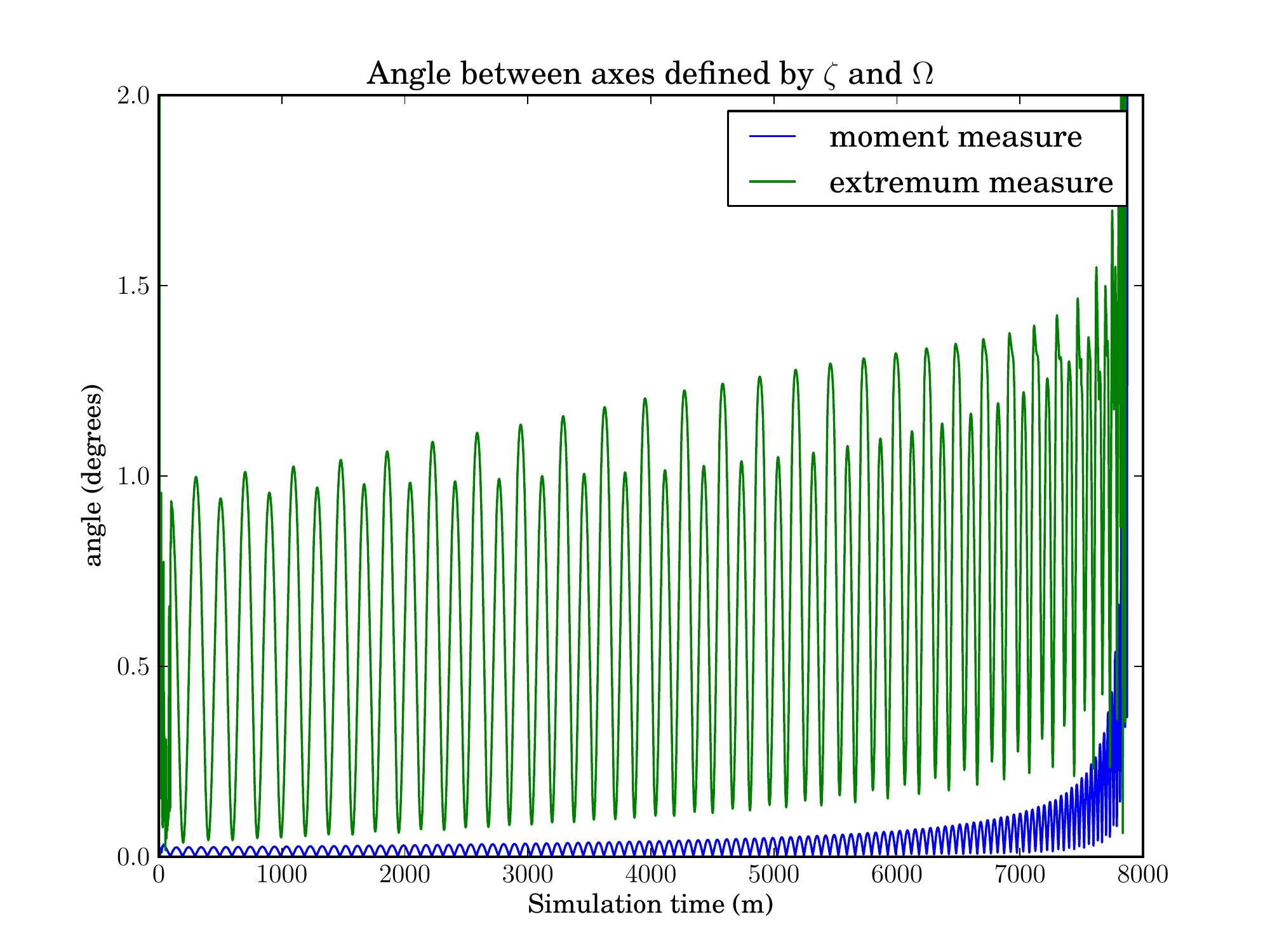}
 \caption{\label{f:AKVVsOmega} Angle (in the Euclidean background space) 
between spin axes defined using $\zeta$ and $\Omega$, either using coordinate 
moments or a line between extrema. The curves are visually identical if 
$\Omega$ is replaced by $B_{ss}$. Interestingly, the two scalars agree to 
within our rough estimate of numerical truncation, so long as the 
axis is defined using coordinate moments of. The 
measures agree far less well when extrema are used. This is 
apparently a consequence of the higher-multipole structure present in $B_{ss}$ 
and $\Omega$, already noted in Fig.~\ref{f:ExtremaVsMoments}.}
\end{figure}

\begin{figure}
 \includegraphics[width=\columnwidth]{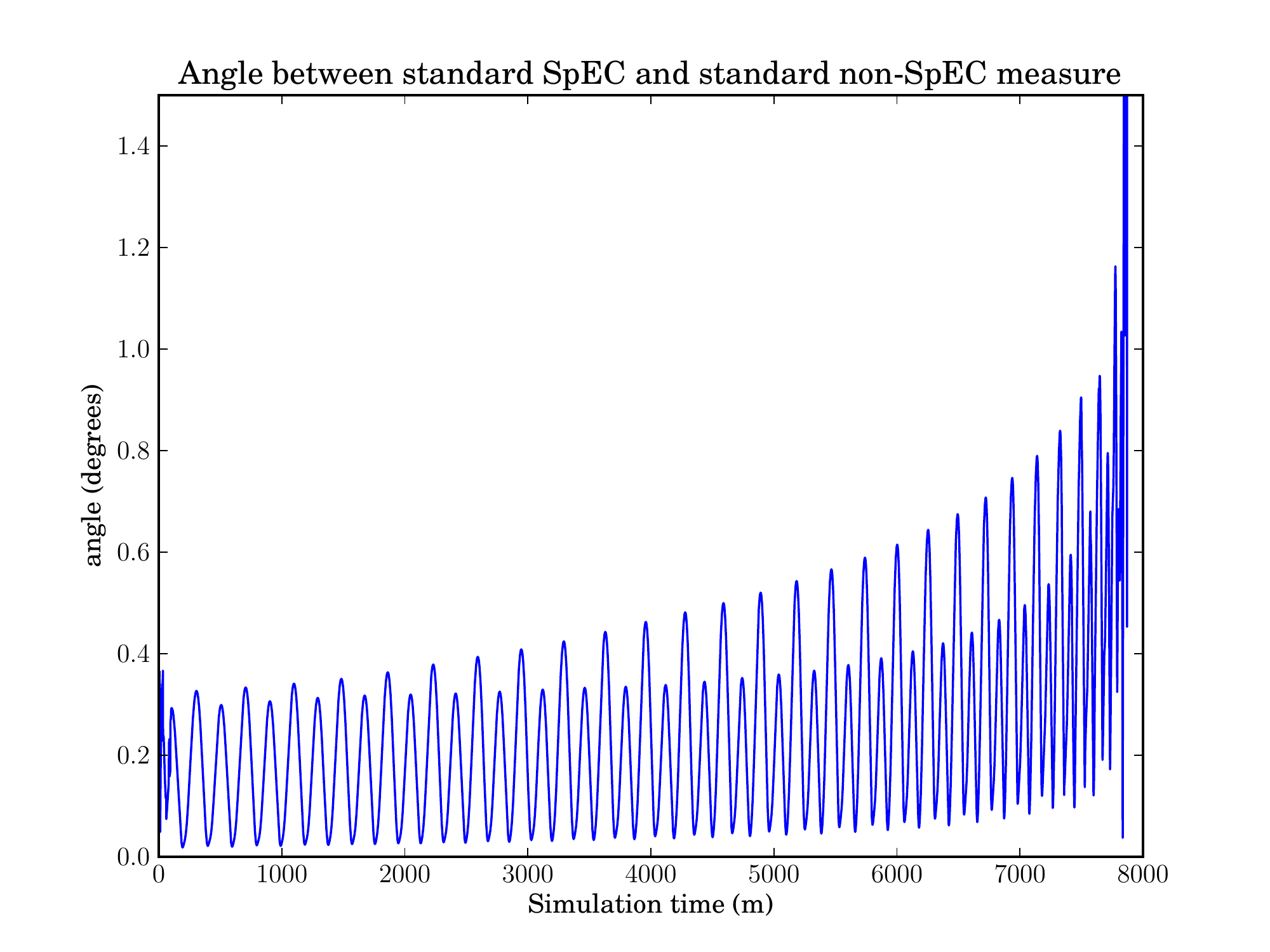}
 \caption{\label{f:SpECVsNonSpEC} Angle (in the Euclidean background space) 
between the default~\SpEC~spin axis and our model of the standard spin axis 
in non-\SpEC~codes.}
\end{figure}

In Figures~\ref{f:ExtremaVsMoments} --~\ref{f:SpECVsNonSpEC}, we summarize the 
discrepancies between these measures. 

Figure~\ref{f:ExtremaVsMoments} shows the difference between an axis determined by the line between extrema and an axis determined by coordinate moments. Curves are shown comparing these measures for the three scalar quantities $\zeta$, $B_{ss}$, and $\Omega$, though the latter two curves overlap to the eye. The fact that the measures differ more for $\Omega$ and for $B_{ss}$ than for $\zeta$ is likely because $\Omega$ and $B_{ss}$ carry higher multipolar structure. Indeed, the $\zeta$ quantity is considered in Ref.~\cite{Owen2009} to define a pure dipole, in a spectral sense, which would be expected to agree reasonably well with a coordinate dipole, for which the moment measure and extremum measure would be expected to agree exactly. 

Figure~\ref{f:BssVsOmega} shows the differences between using $B_{ss}$ and $\Omega$, either via moments or extrema. The agreement is remarkably good, well under our rough estimate of $.02^\circ$ numerical uncertainty. This implies that the pre-merger dynamics are not strong enough to cause $\Omega$ and $B_{ss}$ to substantially differ from one another. 

Figure~\ref{f:AKVVsOmega} shows the differences between a spin axis determined by the symmetry generator $\zeta$ and the normal-bundle curvature $\Omega$, either via coordinate moments or extrema. Here, the moment measures agree much better (differences of order $.05^\circ$) than the extremum measures (differences of order $1^\circ$). The greater variation in extremum measures is again due to the fact that $\Omega$ has more non-dipolar structure than $\zeta$. This non-dipolar structure is largely filtered out from the coordinate moments.

As a final direct comparison, we show the discrepancy between the default measure of spin axis in~\SpEC~(moments of $\Omega$, $\hat \chi _{\Omega m}$) with our model of the standard non-\SpEC~measure (the line between rotation poles, which we model as the line between extrema of $\zeta$, $\hat \chi_{\zeta e}$). The discrepancy is within half a degree for much of the inspiral, implying that both measures would be essentially equally ``good'' for rough measurements of the spin axis and its secular precession, but without further theoretical justification, likely neither is trustworthy for the finer nutation features that motivate the current work. 

\begin{figure}
 \includegraphics[width=\columnwidth]{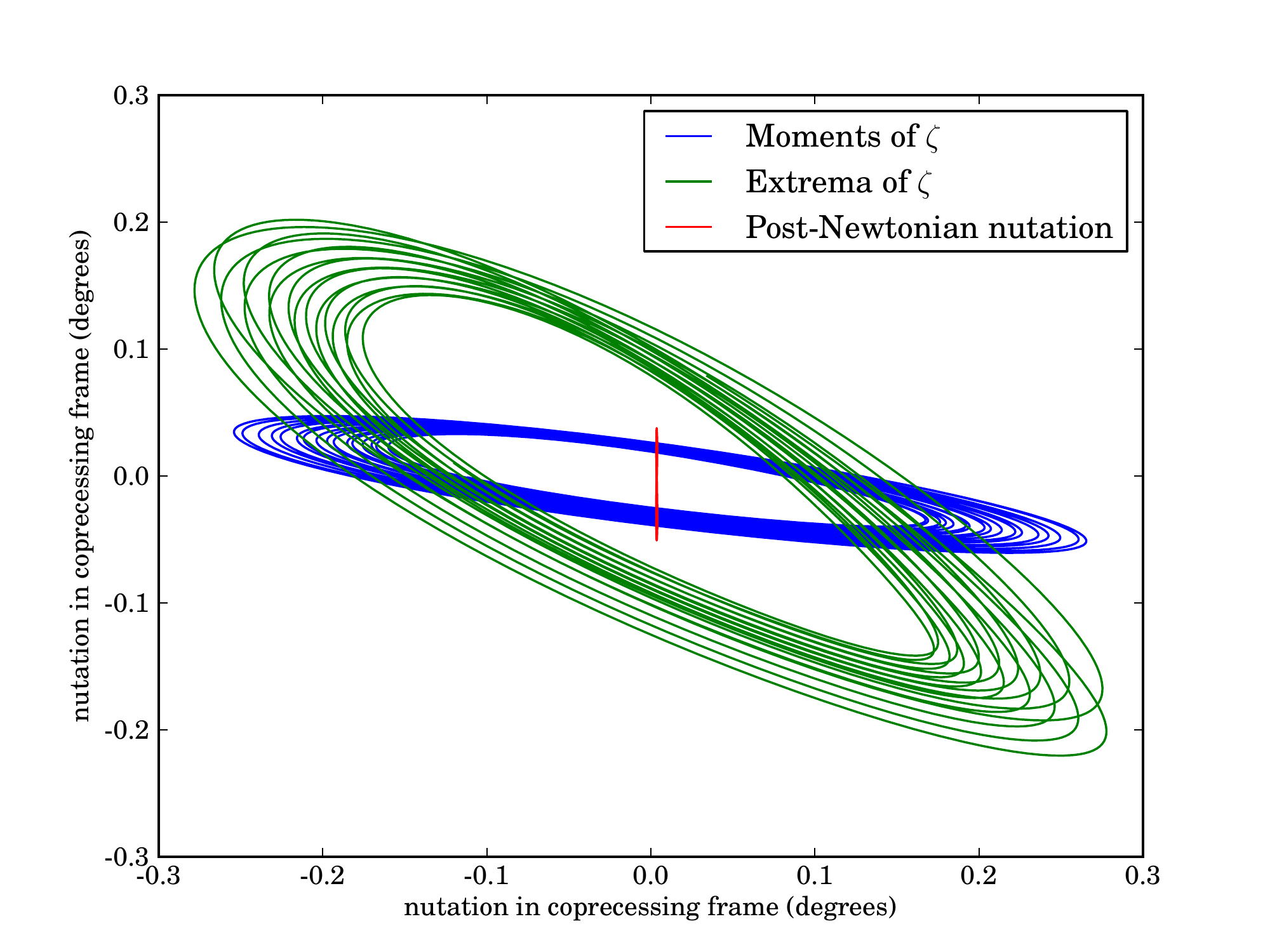}
 \caption{\label{f:ZetaNutation} Nutation of the spin axes $\hat \chi_{\zeta m}$ and $\hat \chi_{\zeta e}$, along with the nutation expected from post-Newtonian theory. The quantity $\vec \chi \cdot \hat e_2$ is plotted on the horizontal axis and $\vec \chi \cdot \hat e_3$ on the vertical axis, where $\hat e_i$ are defined in the text. The expectation from post-Newtonian theory is that the nutation should be represented as purely vertical oscillation on this chart, whereas the measure involving moments of $\zeta$ implies unphysical horizontal nutations greater than the vertical nutations by approximately a factor of five. The measure involving extrema of $\zeta$ (our model of the standard non-\SpEC~axis measure) shares this unphysical horizontal nutations as well as drastically enlarged vertical nutations. 
}
\end{figure}

\begin{figure}
 \includegraphics[width=\columnwidth]{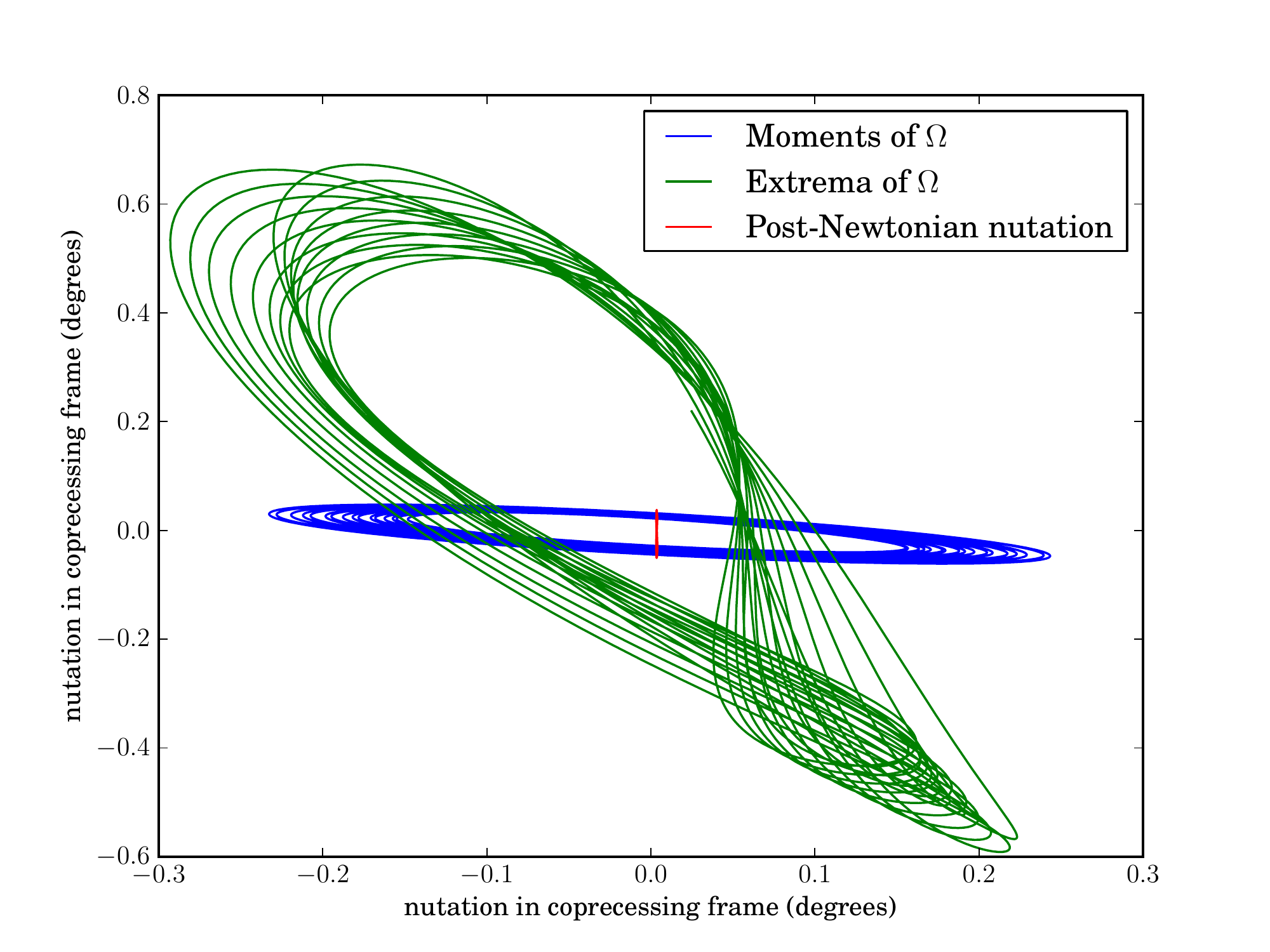}
 \caption{\label{f:OmegaNutation} Nutation of the spin axes $\hat \chi_{\Omega m}$ and $\hat \chi_{\Omega e}$, along with the nutation expected from post-Newtonian theory. The quantity $\vec \chi \cdot \hat e_2$ is plotted on the horizontal axis and $\vec \chi \cdot \hat e_3$ on the vertical axis, where $\hat e_i$ are defined in the text. The quantity $\hat \chi_{\Omega m}$, which is the default measure of spin axis in the~\SpEC~code, shares with $\hat \chi_{\zeta m}$ the large unphysical nutations in the horizontal direction. The nutation of $\hat \chi_{\Omega e}$ is even stranger, due to the higher multipolar structure in $\Omega$. We have also constructed the analogous diagram for $\hat \chi_{B_{ss} m}$ and $\hat \chi_{B_{ss} e}$, confirming that the nutations for that quantity are visually identical to those shown here. 
}
\end{figure}

To further explore the variation in these axis measures, we repeat a procedure from Ref.~\cite{Ossokine2015} that allows us to trace out the finer nutation features without the distraction of the secular precession. The method is based around the idea of a ``coprecessing'' frame. We begin by computing a running average of any given spin axis vector $\hat \chi$ (represented as a vector in the space associated with the flat reference geometry of the $x^i$ coordinates), averaging over a few orbital periods (therefore multiple nutation cycles) to define a slowly, steadily rotating vector $\hat e_1$. We then compute the time derivative of this steadily rotating vector and normalize it to define another basis vector $\hat e_2$. We then complete the triad with a simple cross product: $\hat e_3 = \hat e_1 \times \hat e_2$. The nutation, referring to the variation in spin axis that occurs within the window of time averaging, would be represented by the quantities $\vec \chi \cdot \hat e_2$ and $\vec \chi \cdot \hat e_3$. Figures~\ref{f:ZetaNutation} and~\ref{f:OmegaNutation} apply this technique to plot the nutation about the slowly-varying orbit-averaged precession axis. In both cases we also include the nutation expected from a post-Newtonian calculation, which we compute using equations available in Ref.~\cite{Ossokine2015}. The post-Newtonian calculations tell us to expect purely vertical nutations (that is, nutations purely perpendicular to the direction of the averaged precession of the spin axis) with amplitude of approximately $0.05^\circ$. The moments of $\zeta$, $\Omega$, and $B_{ss}$ all roughly agree with this ``vertical'' nutation expected from post-Newtonian theory, yet they add a horizontal component (that is, {\em along} the direction of the averaged precession of the spin axis) which does not vanish as expected, and indeed exceeds the vertical nutation by approximately a factor of five. The axis measures defined by extrema nutate even more wildly, with a significant increase in the vertical nutation in $\hat \chi_{\zeta e}$ and even more significant variations in $\hat \chi_{\Omega e}$ and $\hat \chi_{B_{ss} e}$, again attributable to the higher multipolar structure of these quantities. One can also note a ``two-leaved'' quality to the curves in Figs.~\ref{f:ZetaNutation} --~\ref{f:OmegaNutation}, particularly on the curves for extremum-based measures (green). We assume this is due to remaining eccentricity in the simulation, with two nutation cycles per orbital cycle, one with slightly greater separation than the other. 

\section{Another look at coordinate rotation generators}
\label{s:newmeasure}

It should not be particularly surprising that the spin axis measures defined in 
Sec.~\ref{s:existingtechniques} behave in a qualitatively different fashion 
than the spin of post-Newtonian theory. Even aside from the subtle questions 
of gauge ambiguity, a much larger issue is that there simply is no reason that 
the axis of approximate horizon symmetry (defined either by poles or by 
moments), or the same axes defined by horizon vorticity, should behave 
dynamically in the same manner as the spin defined in post-Newtonian theory. 
They are simply {\em different} quantities, intuitively expected to agree in 
some vague approximate sense, but not with the kind of precision that is 
available in modern numerical relativity codes. 

\begin{figure}
 \includegraphics[width=\columnwidth]{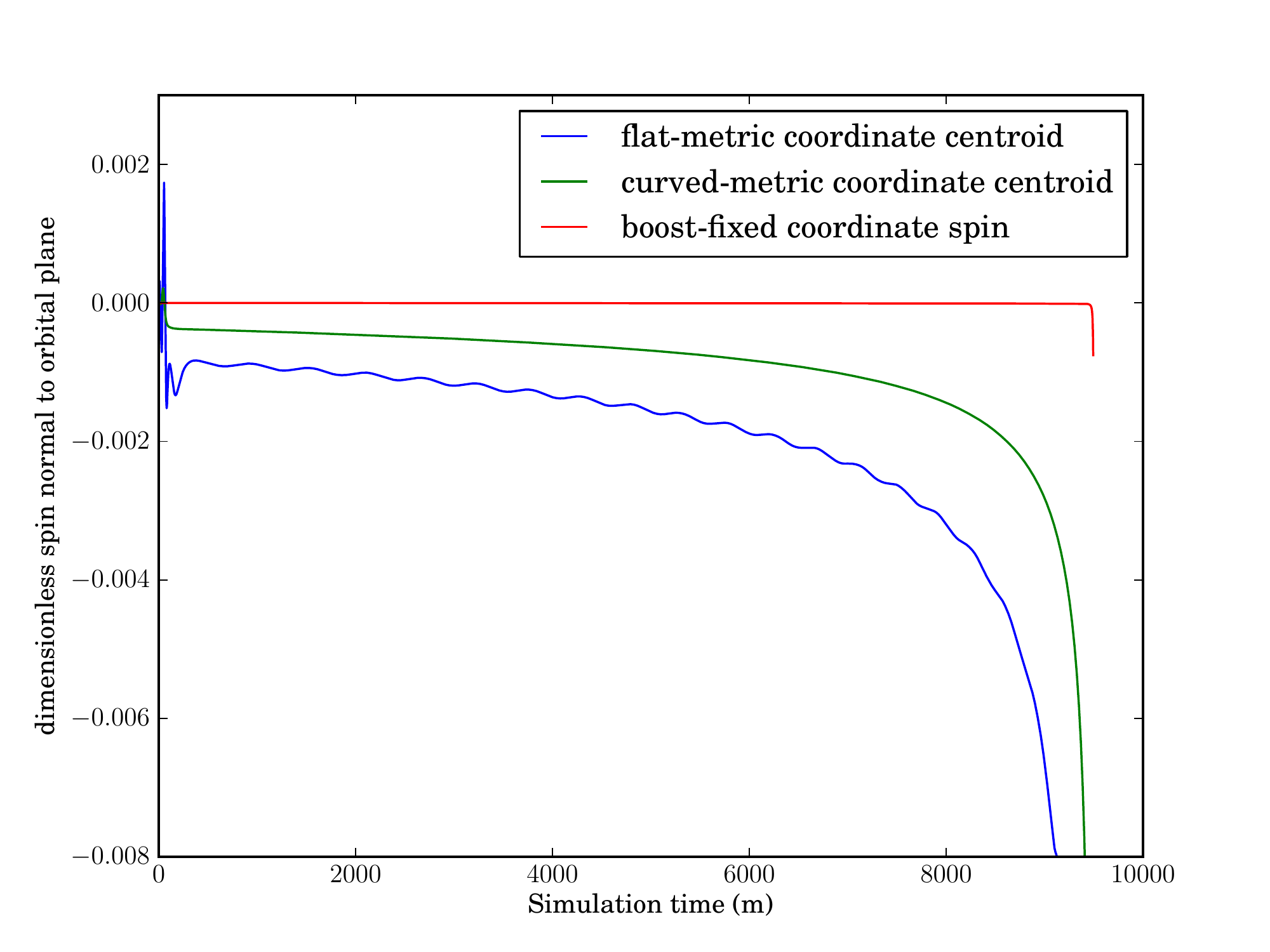}
 \caption{\label{f:WrongWaySpinup} Component of spin, on either black hole, along direction of orbital angular momentum, in a binary of equal-mass initially nonspinning black holes. When the spin is measured using the simple coordinate rotation vectors of Eq.~\eqref{e:CoordinateRotationGenerators1}, the holes spin up in the direction opposite expectations from perturbation theory. This spinup, while small, is well resolved by the code. Moreover, this spinup is strongly influenced by the choice of the coordinate centroid defining the rotation vectors. When the surface angular momentum density $\omega_A$ is ``boost-fixed,'' by setting its gradient-potential $\pi$ (defined in Eq.~\eqref{e:HodgeDecomposition}) to zero, the spin (with either centroid) remains zero to well within the accuracy of numerical truncation throughout the entire inspiral (including a sharp but numerically unresolved spinup after the formation of the common horizon). 
}
\end{figure}

In order to bring the discussion back to concepts directly associated with 
angular momentum, we return to the quasilocal formula in 
Eq.~\eqref{e:CoordSpin}. Again, this spin measure was explored in the early 
days of the~\SpEC~code, but abandoned. It was abandoned in part because it did 
not satisfy the useful technical conditions of centroid invariance or 
boost-gauge invariance, but a more direct reason was the nonphysical behavior 
it exhibited in binaries of small spin. This behavior, in a simple simulation 
of an equal-mass non-spinning inspiral, 
can be seen in Fig.~\ref{f:WrongWaySpinup}. The holes are 
nonspinning according to the well-defined spin magnitude computed 
using approximate Killing 
vectors~\cite{OwenThesis, Cook2007, Lovelace2008}.
 After the initial ringing, 
a small but well-resolved spin in the $-z$ direction arises (that is, in the direction opposite the orbital angular momentum). More troublingly, this nonzero spin grows 
over the course of the inspiral, still in the direction opposite the orbital 
angular momentum. This spinup is opposite (and much stronger than) what would 
be expected from tidal spinup. 

Figure~\ref{f:WrongWaySpinup} includes curves showing two choices of 
centroid. The blue curve uses the centroid 
computed directly from the coordinate shape, assuming {\em no} spatial 
curvature:
\begin{equation}
x_{0, flat}^i = \frac{1}{A_0}\oint_{\cal H} x^i \dA_0,\label{e:FlatCentroid}
\end{equation}
where $dA_0$ refers to the surface area element inferred on the horizon 
2-surface ${\cal H}$ from the flat background geometry. The green curve shows 
the spin computed similarly, but fixing the centroid using the physical 
curved-space geometry:
\begin{equation}
x_{0, curved}^i = \frac{1}{A} \oint_{\cal H} x^i \dA.\label{e:GeometricCentroid}
\end{equation}

Figure~\ref{f:WrongWaySpinup} not only shows the unphysical apparent spinup, it 
also clearly shows that it is strongly dependent on the 
choice of centroid. Thus, by fixing the centroid in a reasonable way, one 
might hope to cure the apparent spinup.

More precisely, if the centroid coordinates are offset by three constants $x_0^i \mapsto x_0^i - \Delta x^i$, then the coordinate rotation vectors defined by Eqs.~\eqref{e:CoordinateRotationGenerators1} or~\eqref{e:CoordinateRotationGenerators2} transform as:
\begin{equation}
\vec \phi_{i} \mapsto \vec \phi_{i} + \eta_{ij}{}^k \Delta x^j \vec \tau_k,
\end{equation}
where $\vec \tau_k$ is the translation-generating vector used in the definition of the rotation generator ($\vec \partial_k$ for $\vec \phi_{i}$, $\delta_{kl} g^{ab}\partial_b x^l$ for $\vec \varphi_{i}$). The result of this transformation on the angular momentum integral, Eq.~\eqref{e:CoordSpin}, is the familiar:
\begin{equation}
J_{i} \mapsto J_{i} + \eta_{ij}{}^k \Delta x^j p_k, \label{e:LinMomCorrection}
\end{equation}
where: 
\begin{equation}
p_k := \frac{1}{8 \pi_0} \oint \vec \omega \cdot \vec \tau_k \hspace{1mm} dA.\label{e:LinearMomentum}
\end{equation}
This quantity is naturally interpreted as a quasilocal {\em linear} momentum, and has been explored numerically in Ref.~\cite{Krishnan:2007pu}. Its significance here lies in the fact that if a hole is in motion, then the centroid ambiguity directly implies an ambiguity in the spin magnitude and axis defined by the components $J_{i}$ in the global background coordinates. This ambiguity (of separating ``spin'' angular momentum from ``orbital'' angular momentum) is familiar from even basic Newtonian physics and post-Newtonian theory. In the nonrelativistic context, the ambiguity is generally fixed by placing the centroid at the center of mass of the moving object. In the relativistic context, however, this strategy fails because the concept of the center of mass is not Lorentz covariant~\cite{Schiff1960}.

\subsection{Hodge decomposition and boost-fixed coordinate spin}
\label{s:Hodge}

The above ambiguity in the black hole centroids is not the only ambiguity that we must 
contend with. Another arises from the fact that the black hole is necessarily 
an extended object. Let us decompose the $\vec \omega$ quantity into two 
scalar potentials on the horizon:
\begin{equation}
\omega^A = D^A \pi + \epsilon^{AB} D_B \varpi. \label{e:HodgeDecomposition}
\end{equation}
If we know the vector $\vec \omega$, then the potentials $\pi$ and $\varpi$ can 
be readily computed through a Poisson equation on the 2-surface:
\begin{align}
D^2 \pi &= D_A \omega^A = \Pi,\label{e:pipoisson}\\
D^2 \varpi &= \epsilon^{AB} D_A \omega_B = \Omega.\label{e:varpipoisson}
\end{align}
Note that because the sources, $\Pi$ and $\Omega$, are pure derivatives, they 
integrate to zero on 
the 2-surface, the necessary condition for them to lie within the image of the 
horizon Laplacian operator, and thus for solutions for solutions $\pi$ and 
$\varpi$ to exist. These potentials are defined only up to a constant, but 
the constant 
is irrelevant in reconstructing $\omega_A$, and hence the spin. (In~\SpEC, 
these potentials can be computed, and the constant is fixed by the condition 
that $\pi$ and $\varpi$ integrate to zero.)

Note that because $\Omega$ is boost-gauge invariant, the potential $\varpi$ 
defined by Eq.~\eqref{e:varpipoisson} is also boost-gauge invariant. The other 
potential, $\pi$, is {\em not} boost-gauge invariant, and for interesting 
physical reasons. Under a boost-gauge transformation, 
$\vec l \mapsto e^a \vec l$, $\vec n \mapsto e^{-a} \vec n$, $\vec \omega$ 
transforms as:
\begin{align}
\omega^A &\mapsto \omega^A - D^A a,\\
D^A \pi + \epsilon^{AB} D_B \varpi &\mapsto D^A \pi + \epsilon^{AB} D_B \varpi - D^A a,
\end{align}
and hence we can infer that, up to an additive constant, 
$\pi \mapsto \pi - a$. In the context of the rotation generators $\vec \varphi_{i}$ defined in Eq.~\eqref{e:CoordinateRotationGenerators2} we will find that $\pi$ is directly related to the quasilocal {\em linear} momentum, so it is no surprise that it is boost-dependent.

Now, return to the quasilocal angular momentum formula in Eq.~\eqref{e:CoordSpin} (suppressing global coordinate indices $i, j, k, ...$ for simplicity). Substitute the above
scalar decomposition, and integrate by parts:
\begin{align}
J &= \frac{1}{8 \pi_0} \oint_{\cal H} \left( D^A \pi + \epsilon^{AB} D_B \varpi  \right) \phi_A \dA,\\
8 \pi_0 J &= -\oint_{\cal H} \pi \hspace{1mm} \left(D^A \phi_A\right) \dA + \oint_{\cal H} \varpi \left(\epsilon^{AB} D_A \phi_B\right) \dA. \label{e:Jdecomposed}
\end{align}
The first integral in this final expression is not boost-gauge invariant, while 
the second integral is (note that the rotation generator $\vec \phi$ is taken 
to be projected tangent to the 2-surface, and hence is manifestly boost-gauge 
invariant). We could choose to 
fix boost gauge with the condition $\pi = 0$, which is always accessible given 
the transformation law for $\pi$. This fixing of boost-gauge was 
employed by Korzynski~\cite{Korzynski2007} in a different but related approach 
to quasilocal black hole spin (involving conformal Killing vectors on the 
horizon, rather than the global coordinate rotations considered here). 
But there is another way to think about the condition in the current context. 

The offending boost-dependent term in Eq.~\eqref{e:Jdecomposed} also involves the quantity
$\nabla_A \phi^A$, the divergence of the rotation generator. If $\vec \phi$ 
were a true Killing vector on the 2-surface, then this divergence would vanish 
simply due to the trace of Killing's equation. 

As a simple motivating case, consider a metrically-round 2-sphere embedded in 
truly Euclidean 3-space. The Killing vector fields that generate rotations 
about the center of this sphere are tangent to it and thus also 
Killing vectors of the 
surface. One can show that the vector fields that generate rotations about a 
point {\em offset} from the center of the sphere, when projected down to 
the surface, have nonzero surface divergence. Furthermore this surface 
divergence is linear in the offset vector. Hence, in 
Euclidean space, one can fix the centroid ambiguity by insisting that the 
rotation generator has zero divergence.

On an arbitrary 2-surface, in curved or flat geometry, it is no longer true that 
simple translation of background-coordinate rotation generators can always 
render the projected field divergence-free. However one can 
always project out the ``divergence-free part'' of $\vec \phi$, using 
a Hodge decomposition analogous to that already employed for $\vec \omega$:
\begin{align}
\phi_A &= D_A \alpha + \epsilon_A{}^B D_B \beta\\
\beta &= D^{-2} \left(\epsilon^{AB} D_A \phi_B\right)\\
\phi_A^{DF} &= \epsilon_A{}^B D_B \beta
\end{align}
Such a projection might be considered a ``generalized'' translation of the 
coordinate rotation generator.

Either viewpoint (fixing boost-gauge to render $\pi = 0$, or deforming the 
rotation generator to remove its divergence while preserving its curl) leads 
to the following very simple gauge-fixed coordinate spin vector:
\begin{equation}
J_{bf}^i := \frac{1}{8 \pi_0} \oint_{\cal H} \varpi \left( \epsilon^{AB} D_A \phi_B^i \right)\dA,\label{e:BoostFixedCoordSpin}
\end{equation}
where $\vec \phi^i$, for $i=1,2,3$, are the three coordinate rotation generators. We refer to this quantity as the ``boost-fixed'' coordinate spin vector. It should be noted that at this point we are being agnostic about the definition of the coordinate rotation vector. This concept of ``boost-fixing'' the coordinate spin applies either for the generators $\vec \phi_{i}$ defined in Eq.~\eqref{e:CoordinateRotationGenerators1}, or $\vec \varphi_{i}$ defined in Eq.~\eqref{e:CoordinateRotationGenerators2}.

\subsection{Behavior under changes of coordinate centroid}
\label{s:Translation}

As noted in Eq.~\eqref{e:LinMomCorrection}, when the coordinate centroid is offset, our spin measures can change because ``orbital'' angular momentum becomes reclassified as ``spin'' and vice versa. As mentioned in Sec.~\ref{s:Hodge}, this centroid ambiguity is entangled with the boost-gauge ambiguity because translation of the coordinate centroid is associated (at least in simple cases) with changing the ``pure gradient'' part of the rotation generator, which due to parity considerations only interacts in the spin calculation with the boost-dependent potential $\pi$. Indeed, a very deep parallel exists when the $\vec \varphi_i$ rotation generators are used. 

In Eq.~\eqref{e:Jdecomposed}, we see that the boost-gauge dependent potential $\pi$ is integrated with the surface divergence of the rotation generator, while the boost-gauge {\em independent} potential $\varpi$ is integrated against the curl of the rotation generator. In Appendix~\ref{a:TheTwoGenerators}, we show that the $\vec \varphi_i$ rotation generators, which are constructed using translation {\em one-forms} ${\bf d} x^i$, have a surface curl that is independent of the coordinate centroid (Eq.~\eqref{e:CurlFormBasedGen}) and a surface divergence that is linear in it (Eq.~\eqref{e:DivFormBasedGen}). The term that is boost-gauge dependent is also the term that is centroid dependent, and if boost-gauge is fixed by setting the potential $\pi$ to zero, then the centroid dependence disappears completely. 

One way to think of this surprising coupling of boost dependence with centroid dependence is that the condition $\pi = 0$ amounts to measuring the spin in a comoving frame. Translating the coordinate centroid affects the spin through a term proportional to the ``quasilocal linear momentum'' defined in Eq.~\eqref{e:LinearMomentum}. But in this case the coordinate translation generator is ${\bf d} x^k$ and the momentum is:
\begin{eqnarray}
8 \pi_0 p^k &=& \oint \omega^A \partial_A x^k dA,\\
&=& -\oint \left(D_A \omega^A\right) x^k dA,\\
&=& -\oint \left(D^2 \pi\right) x^k dA. \label{e:FormBasedLinMom}
\end{eqnarray}
Fixing boost gauge such that $\pi = 0$ means that in this boost gauge the linear momentum vanishes, and hence displacements cannot affect the spin. The important technical consequence of this is that the ``boost-fixed'' spin measure of Eq.~\eqref{e:BoostFixedCoordSpin}, when computed using the $\vec \varphi_i$ rotation generators, is {\em also} centroid independent. 

This remarkably satisfying connection between boost and centroid freedom unfortunately does not carry over to the case when the rotation generators are $\vec \phi_i$, constructed from vectorial translation generators $\vec \partial_i$. This fact is all the more disappointing because, as we will see in the next subsection, the $\vec \phi_i$ rotation generators give much better agreement with post-Newtonian nutation than the $\vec \varphi_i$ generators. 

In Eqs.~\eqref{e:CurlVecGen} and~\eqref{e:NablaTau}, we see that the surface curl of $\vec \phi_i$ includes a term proportional to $x_0^j$ and the physical metric's Christoffel symbols of the first kind $\Gamma_{[da]c}$, antisymmetrized on the {\em first} pair of indices (as opposed to the pair associated with the torsion tensor that we take to vanish). This combination of Christoffel symbols does not vanish in general, and is proportional to partial derivatives of the components of the physical metric. 

The fact that simply boost-fixing to $\pi = 0$ does not provide a centroid-invariant spin measure for the $\vec \phi_i$ generators can also be seen from the linear momentum of Eq.~\eqref{e:LinearMomentum}. Hodge-decomposing the $\vec \omega$ field again, $p_k$ becomes:
\begin{equation}
8 \pi_0 p_k = - \oint \varpi \epsilon_A{}^B D_B \tau^A_k dA - \oint \pi D_A \tau^A_k dA.
\end{equation}
The first integral on the right-hand side here does not necessarily vanish. The coordinate-translation-generating vector fields $\vec \tau_k = \vec \partial_k$ need not be curl-free, even when projected down to the 2-surface. Therefore the odd-parity momentum potential $\varpi$, which {\em cannot} be altered by the choice of boost gauge, contributes to the linear momentum. Therefore setting $\pi = 0$ does not make the linear momentum vanish, and therefore does not make the spin measure centroid-independent. 

This complication does not mean that it's impossible to choose a boost gauge in which $p_k = 0$. In fact, one can easily enforce this by choosing some set of three or more linearly independent basis functions on the sphere, representing the boost-gauge transformation function $a(\theta, \phi)$ in this basis, and solving a matrix problem to set the transformed $p_k$ to zero. This kind of boost-fixing would render the spin measure centroid-invariant, however it seems needlessly elaborate, and worse, we see no physical motivation for choosing the particular basis functions used on the sphere. This issue could warrant deeper study in the future, but we will ignore it for the remainder of this paper. Instead, we will continue to use the words ``boost-fixed'' to refer to a boost gauge in which $\pi = 0$, even though in the context of the rotation generators $\vec \phi_i$, constructed from translation generators $\vec \partial_k$, this condition does not render the quasilocal linear momentum zero, nor does it make the boost-fixed spin measure centroid invariant.

\subsection{Associated ambiguity of the spin axis}

We began this section with a demonstration that the simple ``coordinate spin'' 
vector of Eq.~\eqref{e:CoordSpin} is at least slightly dependent on the choice 
of centroid for the coordinate rotation generator. This dependence is directly 
analogous to a translation ambiguity that exists even in Newtonian physics, 
and post-Newtonian theory as well, in which the spin 
``vector'' is defined only after a certain ``spin supplementary condition'' 
has been imposed~\cite{Schiff1960}. It is extremely tempting to 
generalize the standard spin supplementary conditions of post-Newtonian 
theory to the context of full numerical relativity. We intend to pursue this 
path in future research, however such a generalization would necessarily 
involve quasilocal {\em energy}, a concept that has been very well studied 
(the approach of Brown and York~\cite{BrownYork1993}, in particular, is 
directly related to the angular momentum arguments treated here). However 
quasilocal energy is famously a much more subtle concept than the simple 
angular momentum constructions employed here, so we must consider this 
extension beyond the scope of the current work.

Instead, here we simply view the centroid ambiguity as another inherent 
ambiguity of the coordinate spin angular momentum. This ambiguity can be fixed 
by choosing particular centroids such as those defined in Eqs.~\eqref{e:FlatCentroid} --~\eqref{e:GeometricCentroid}, or by employing the boost-invariant measure defined in Eq.~\eqref{e:BoostFixedCoordSpin}, as long as the rotation generators $\vec \varphi_i$ are used.

\begin{figure}
  \includegraphics[width=\columnwidth]{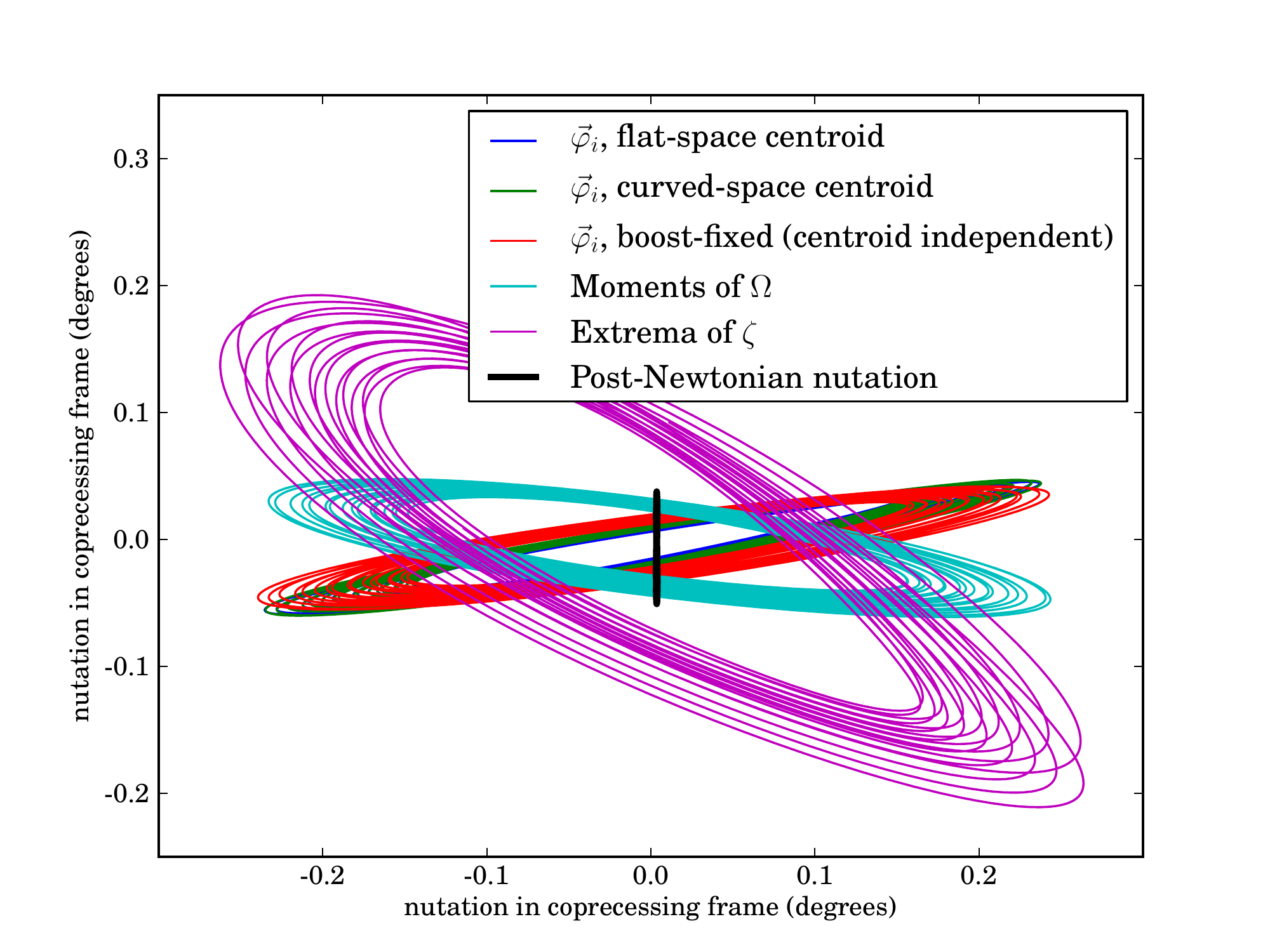}
  \caption{\label{f:FormBasedGenNutationTrack}
    Nutation in a coprecessing frame using the coordinate rotation generators $\vec \varphi_i$, which are constructed from translations defined by the one-form ${\bm \kappa}^i = {\bf d}x^i$. This choice of rotation vector is very nicely behaved, mathematically speaking, however it does not appear to improve upon the spurious nutations along the direction of secular precession, seen also with the spin measures using moments of $\Omega$ ({\tt SpEC} standard), or extrema of zeta (our model of the common standard outside of the {\tt SpEC} code). The ``boost fixing'' procedure described around Eq.~\eqref{e:BoostFixedCoordSpin} does not appear to provide an obvious improvement over the standard coordinate spin calculation, at least with regard to agreement with the precession track of an analogous post-Newtonian calculation, denoted by the heavy black line. All three of our $\vec \varphi_i$-related coordinate spin measures seem qualitatively on par with the {\tt SpEC}-standard measure, though intriguingly the phase lag between horizontal nutations and vertical nutations is different for the coordinate spin measures than for the moments of $\Omega$.
}
\end{figure}

In Fig.~\ref{f:FormBasedGenNutationTrack}, we visualize the nutation computed with three kinds of coordinate spin measure which use the rotation vectors $\vec \varphi_i$ computed from the one-forms $\bm{\varphi}_i := \epsilon_{ijk} \left(x^j - x_0^j\right) {\bf d} x^k$. Specifically, we have fixed $x_0^j$ using the ``flat-space centroid'' (Eq.~\eqref{e:FlatCentroid}) condition, the ``curved-space centroid'' (Eq.~\eqref{e:GeometricCentroid}) condition, and the ``boost-fixed'' variant of the spin measure (Eq.~\eqref{e:BoostFixedCoordSpin}), which we have found to be independent of the choice of $x_0^j$ when the rotation vectors $\vec \varphi_i$ are used. Despite the nice mathematical features of the $\vec \varphi_i$ rotation vectors, it is clear that calculating spin using these rotation generators does not provide any qualitative improvement over the standard \SpEC~measure which uses coordinate moments of the $\Omega$ scalar. Intriguingly, the spurious horizontal nutations are approximately equal using all the measures shown (though the phase difference between horizontal and vertical nutations is different than for the measures involving $\Omega$ or $\zeta$). The similarity of the horizontal nutations in all of these cases might imply that they all arise from the same source, which we assume is the bulging of the horizon's coordinate shape as the spatial separation vector changes relative to the spin directions. It should also be noted that the boost-fixing procedure, while it does fix boost gauge and might be taken to slightly reduce spurious vertical nutations (which can be compared with the post-Newtonian nutation track drawn heavily in black), it does not appear to have any significant effect on the spurious horizontal nutation. 

\begin{figure}
  \includegraphics[width=\columnwidth]{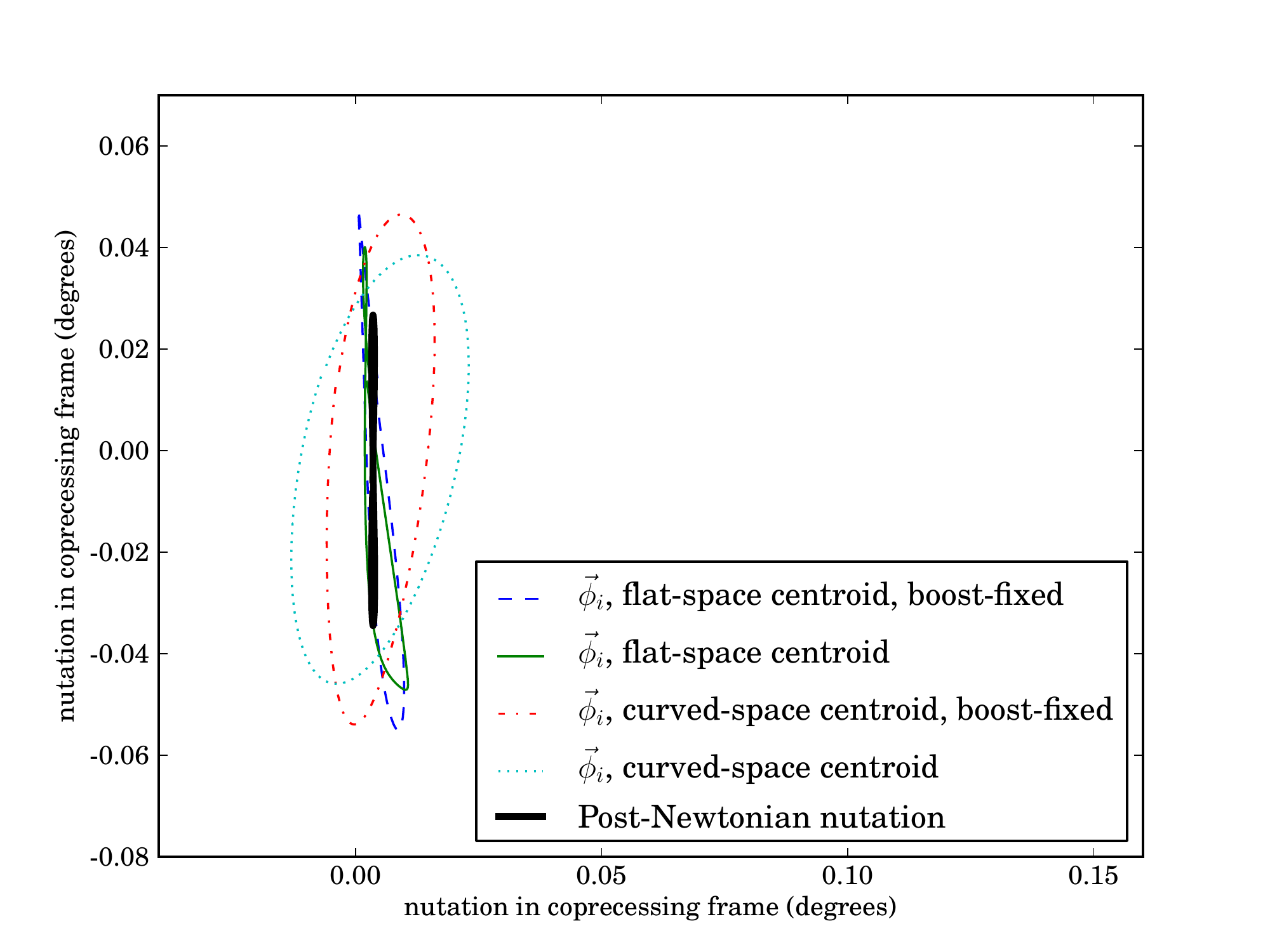}
  \caption{\label{f:VectorBasedGenNutationTrack}
    Nutation in a coprecessing frame using coordinate rotation generators $\vec \phi_i$, which are constructed from translations defined by the standard coordinate translation vectors $\vec \tau_i = \vec \partial_i$. To keep the graph from getting cluttered, we have chosen a single cycle of precession near the beginning of the inspiral (after junk radiation has dissipated), though later cycles are qualitatively similar, as in other figures of the nutation track above. All four of the nutation tracks shown have dramatically better qualitative agreement with the post-Newtonian expectation (heavy black curve), with the spurious horizontal nutations reduced by an order of magnitude compared with the other measures in this paper. The remaining horizontal nutations are now roughly equal to the discrepancy of the vertical nutations with the post-Newtonian track, and hence we suspect they may all be due to remaining inaccuracies of the numerical relativity simulation such as numerical truncation error (which we've estimated to be around $.02^\circ$), residual orbital eccentricity (the post-Newtonian calculation assumes a non-eccentric orbit), or possibly genuine nonlinear behavior in the numerical relativity simulation.
}
\end{figure}

In Fig.~\ref{f:VectorBasedGenNutationTrack}, we see a drastic improvement that occurs with these measures when the rotation vectors are changed from $\vec \varphi_i$ to $\vec \phi_i := \epsilon_{ij}{}^k \left(x^j-x_0^j\right) \vec \partial_k$. The spurious horizontal nutations have decreased by an order of magnitude, down to a level where they are comparable with the deviations in the vertical nutation from post-Newtonian expectations. Again, the boost-fixing procedure does not seem to make a notable qualitative difference.

It may be tempting to argue that $\vec \phi_i$ would inevitably behave better than $\vec \varphi_i$, because the translation vectors $\vec \partial_k$ from which the former are constructed are the ``true'' translation generators. In a geometrical sense, though, this is not completely obvious. Near the horizon of a black hole, {\em neither} the $\vec \varphi_i$ nor the $\vec \phi_i$ will satisfy Killing's equation, and in the limit of flat geometry, {\em both} sets will satisfy it (assuming both the $x^i$ and the $x^a$ coincide with the Cartesian system in that limit). One distinction to be noted is that the three $\vec \partial_i$ vectors constitute a mutually commuting triple, whereas the three vector fields $\vec \kappa_i$ with components $\kappa_i^a = g^{ab}\delta_{ij} \partial_b x^j$ do not necessarily commute. 

As a final demonstration of the improved behavior of these coordinate spin axes, we repeat the kind of fit to post-Newtonian theory carried out in Ref.~\cite{Ossokine2015}. We solve the same post-Newtonian equations as in that research, though we note that while this reference includes an extremely convenient compilation of orbital, spin-orbit, and spin-spin terms from a variety of sources, it currently includes a few typographical errors. The quantity $\gamma$ should read:
\begin{widetext}
\begin{align}
\gamma = & x \left\{ 1 + \frac{3-\nu}{3} x + \frac{3 \delta \sigma_l + 5 s_l}{3} x^{3/2} + \frac{12 - 65\nu}{12} x^2 + \left(\frac{30+8\nu}{9}s_l + 2 \sigma_l \delta\right)x^{5/2}\right.\nonumber\\
& + \left[1+ \nu\left(-\frac{2203}{2520} - \frac{41\pi^2}{192}\right) + \frac{229\nu^2}{36} + \frac{\nu^3}{81}\right]x^3 + \left(\frac{60-127\nu-72\nu^2}{12}s_l + \frac{18-61\nu-16\nu^2}{6}\sigma_l \delta\right)x^{7/2}\nonumber\\
&+\left. x^2\left(\vec s_0^2 - 3(\vec s_0 \cdot \vec \ell)^2\right)\right\},
\end{align}
and the quantity $b_7$ should read:
\begin{align}
b_7 = &\left(\frac{476645}{6804}+\frac{6172}{189}\nu-\frac{2810}{27}\nu^2\right)s_l + \left(\frac{9535}{336} + \frac{1849}{126}\nu-\frac{1501}{36}\nu^2\right) \delta \sigma_l\nonumber\\
&+ \left(-\frac{16285}{504} + \frac{214745}{1728}\nu + \frac{193385}{3024}\nu^2\right)\pi.
\end{align}
\end{widetext}

We keep all terms in these expressions and the rest of the post-Newtonian expressions given in Ref.~\cite{Ossokine2015}, and handle the evolution of the $x$ parameter using the simple TaylorT1 approximant. We match by minimizing the same integral as in that paper, an integral of:
\begin{equation}
{\cal S} := \langle(\angle L)^2\rangle + \langle(\angle \chi_1)^2\rangle + \langle(\Delta \Omega)^2\rangle,
\end{equation}
where $\angle L$ is the angle (in radians) between the orbital angular momentum axis of the PN solution and that computed from coordinate trajectory data in the numerical relativity solution; $\angle \chi_1$ is the angle between PN and NR spin axes (on the larger hole -- the smaller hole is nonspinning), and $\Delta \Omega := (\Omega_{PN} - \Omega_{NR})/\Omega_{NR}$, where $\Omega_{PN} = x^{3/2}/m$ and $\Omega_{NR}$ is the angular velocity of the NR solution, again computed using coordinate trajectory data. The angled brackets $\langle . \rangle$ refer to a coordinate-time integration, in this case carried out from $t=4000m$ to $t=5500m$, in rough agreement with the window used in Ref.~\cite{Ossokine2015}. 

\begin{figure}
 \includegraphics[width=\columnwidth]{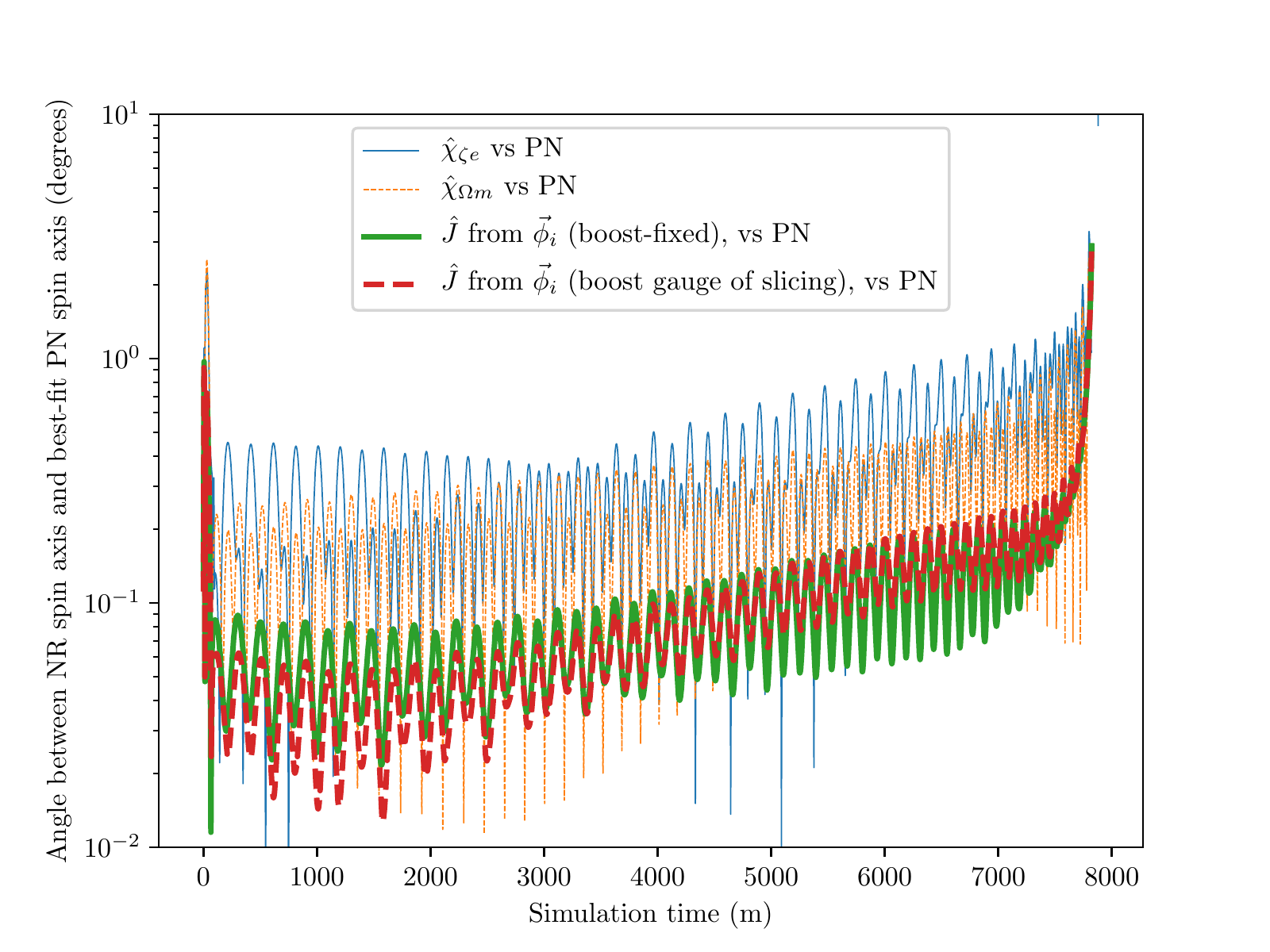}
 \caption{\label{f:FitsToPn} 
Angle between post-Newtonian spin axis and four possible numerical-relativity spin axes. Fitting the post-Newtonian initial data to minimize the quantity ${\cal S}$ defined in the text, where the integration window of the fit extends from $t=4000m$ to $t=5500m$. The uppermost curve (light, solid) shows the discrepancy with $\hat \chi_{\zeta e}$, our model of a standard axis measure defined by a line between poles of the approximate horizon symmetry. The intermediate curve (light, dashed) represents the standard axis measure in the~\SpEC~code, $\hat \chi_{\Omega m}$, which also differs from the best-fit post-Newtonian spin axis by approximately a half degree throughout the inspiral, due to the wild horizontal oscillations visible in Fig.~\ref{f:OmegaNutation}. The bottom curves show the best fit to spin measures based on the quasilocal angular momentum formula using $\vec \phi_i$ as rotation generators. The heavy, solid curve employs the fixing of boost gauge described in the text, while the heavy dashed curve calculates the spin in the default boost-gauge of the simulation, adapted to the spacetime slicing. These measures also use the ``flat space centroid'' condition defined in Eq.~\eqref{e:FlatCentroid}, but the curved-space centroid gives similar results.  As in the nutation tracks of Fig.~\ref{f:VectorBasedGenNutationTrack}, agreement with post-Newtonian expectations is better for the angular-momentum-based measures by approximately an order of magnitude. 
}
\end{figure}

The discrepancy of four numerical-relativity axis measures, versus best-fit post-Newtonian values, is shown in Fig.~\ref{f:FitsToPn}. The spin measure defined in~\eqref{e:BoostFixedCoordSpin}, and the standard angular momentum with coordinate rotation generators $\vec \phi_i$ (using the ``flat-space centroid'' defined by Eq.~\eqref{e:FlatCentroid}), improves on the two standard axis measures in modern Numerical Relativity literature by an order of magnitude. We attribute the failure to find better agreement in the bottom curve, particularly within the matching window itself, to remaining eccentricity in the numerical relativity simulation.

\section{Discussion}
\label{s:discussion}

The research presented here was motivated by two specific goals. The first, more modest, was to simply present formally the default definition of spin axis currently employed in the~\SpEC~code, how it relates to the other current standard and other similarly well-motivated measures. The second goal was to explain and resolve the unphysical nutation features discovered in Ref.~\cite{Ossokine2015}, which clouded otherwise very strong agreement of \SpEC~simulations with post-Newtonian results. 

On the first point, we have formally defined, in Eq.~\eqref{e:SpECDefaultMeasure}, the axis measure $\hat \chi_{\Omega m}$ that defines the spin axis in all currently published~\SpEC~results, and which indexes the simulations catalogued at~\cite{SXSCatalog}. Specifically this measure defines the spin axis through coordinate moments of a scalar quantity $\Omega$ defined on any 2-surface in spacetime, mathematically a scalar curvature of the normal bundle of the embedding of the 2-surface in spacetime. Though it may seem mathematically obscure, this quantity has a long history in~\SpEC, and in the general formalism of quasilocal spin. It is also used in~\SpEC's standard measure of spin magnitude and higher current multipoles~\cite{OwenThesis, Cook2007, Lovelace2008, Owen2009}, as well as a measure of horizon {\em extremality}~\cite{BoothFairhurst:2008, Lovelace:2014}. It is also very closely related (though not mathematically identical) to the normal-normal component of the magnetic part of the spacetime Weyl tensor, $B_{ss}$, which is related to differential frame-dragging at the horizon and referred to as {\em horizon vorticity} in~\cite{OwenEtAl:2011, Nichols:2011pu, Zimmerman2011, ZhangPRD2, VortexPRD3}. While $\Omega$ and $B_{ss}$ are not mathematically equal, we have shown that for inferring spin axis their difference is practically negligible at least in the case considered here. Though the quantity $\Omega$ is intricately related to black hole spin, the use of its coordinate moments as a measure of spin axis was largely an ad-hoc practical decision, with little theoretical justification. 

The other standard measure of spin axis in the current numerical relativity literature is the Euclidean line connecting the poles of an approximate symmetry of the horizon. The approximate symmetry in this context is a vector field computed over the horizon's coordinate grid via integration of Killing transport equations. A direct implementation of this technique would be difficult in~\SpEC, so we have instead considered the line connecting poles of~\SpEC's standard approximate Killing vector, defined via a scalar potential $\zeta$ that satisfies a certain eigenproblem derived from the squared residual of Killing's equation~\cite{OwenThesis, Cook2007, Lovelace2008}. This axis measure, $\hat \chi_{\zeta e}$, while not precisely the same as the measure used in other codes, is directly analogous and so we treat it as a model of the standard measure in other codes.

While we have not carried out a systematic comparison of these standard spin measures, in Fig.~\ref{f:SpECVsNonSpEC} we have given the first direct comparison of these spin measures in a nontrivially precessing binary black hole simulation, showing that over the course of the inspiral these measures agree to within approximately a degree over the course of the inspiral. This is an encouraging sanity check for general purposes, implying that the distinction is likely unimportant for calculations requiring only a coarse measure of the spin axis. At the same time, though, it implies that the features of spin nutation, which in this case oscillate by significantly less than a degree, cannot be expected to be measured accurately by these ad-hoc prescriptions. We have also explored other similar variants, such as using a line between extrema of $\Omega$ and $B_{ss}$, or coordinate moments of the symmetry potential $\zeta$, finding that extrema of $\Omega$ and $B_{ss}$ vary even more wildly, likely due to their higher multipolar structure. 

With regard to the second goal of this paper --- explaining and mitigating the unphysical nutations discovered in Ref.~\cite{Ossokine2015} --- the previous considerations have motivated our view that these nutations were due to the corruption of these ad-hoc axis measures by the tidal structure, present in black hole inspiral, that these ad-hoc measures nonetheless completely ignore. 

To probe spin axis in a more dynamically-meaningful way, we have returned to the quasilocal spin angular momentum measure defined in Eq.~\eqref{e:CoordSpin}. Employing this formula in this context requires us to make some background-dependent decision of what to use for the rotation generators $\vec \phi_i$. The simple application of coordinate rotation generators, while not mathematically elegant, is no less geometrically meaningful than the ad-hoc measures currently employed. There are, however, two reasons to hesitate before using Eq.~\eqref{e:CoordSpin} with coordinate rotation generators, reasons that are not shared by the ad-hoc measures in current use: the resulting spin axis will not necessarily be boost-gauge invariant, and more troublingly, it will depend on the centroid used to define the coordinate rotation. This latter concern, however, suffuses the treatment of spin even in Newtonian and post-Newtonian physics, in which a {\em spin-supplementary condition} must be imposed to specify what precisely one means by the spin vector. With coordinate rotation vectors as defined in Eq.~\eqref{e:CoordinateRotationGenerators1}, the quasilocal spin vector can be written in a form directly analogous to the {\em spin tensor} used in post-Newtonian theory:
\begin{align}
S_i = \frac{1}{8 \pi_0} \oint_{\cal H} \eta_{ijk} \left(x^j - x_0^j\right)\omega^k \dA\\
S^{jk} = \frac{1}{8 \pi_0} \oint_{\cal H} \left(x^{[j} - x_0^{[j} \right) \omega^{k]} \dA,
\end{align}
where $\omega^k := \delta^{kl} e_l^A \omega_A$. In post-Newtonian theory, the spin-supplementary condition, which fixes the worldline of the centroid $x_0^j(t)$, is generally stated as a condition on the spacetime spin tensor, such as $S^{\mu \nu} u_\nu = 0$ where $\vec u$ is, say, the 4-velocity of the spinning body. The appearance of a spin tensor raises the tantalizing possibility of enforcing spin-supplementary conditions of post-Newtonian theory in the numerical relativity context. Unfortunately to do so we would also need to define the space-time components $S^{0i}$ of our numerical relativity spin tensor, and to do that would require implementation of a quasilocal energy measure, which we intend to pursue in future work. 

Instead, we have simply treated the ambiguity of the centroid $x_0^j$ as precisely that, an ambiguity of the formalism. And we have fixed this ambiguity in two ways, by setting $x_0^j$ to the {\em coordinate} center of the horizon, Eq.~\eqref{e:FlatCentroid}, and to the {\em geometric} center of the horizon, Eq.~\eqref{e:GeometricCentroid}. This fixing of coordinate centroid can be seen as analogous to the setting of a spin supplementary condition in PN calculations, but it is frustrating that the two formalisms do not have a common mathematical language.

The measure using coordinate rotation vectors, and either centroid, also suffers from the boost-gauge ambiguity. To deal with this, we have employed a Hodge decomposition of the quasilocal angular momentum density $\omega_A$ to distill its boost-invariant component $\epsilon_A{}^B D_B \varpi$ (in a manner similar to Korzynski~\cite{Korzynski2007}), defining a coordinate spin measure $\vec J_{bf}$, Eq.~\eqref{e:BoostFixedCoordSpin}, that is boost invariant. This boost-gauge fixing did not have a significant impact on the nutation features of our precessing simulation, however it does remove smaller spurious spin features in cases of weakly-spinning holes, as seen in Fig.~\ref{f:WrongWaySpinup}. 

When spin is computed using the coordinate rotation generators $\vec \phi_i$, with or without the boost-fixing trick, and with either of the coordinate centroids employed here (Fig~\ref{f:VectorBasedGenNutationTrack}), the nutation agrees with Post-Newtonian expectations to within our estimated numerical uncertainty of $.02^\circ$, though the nutation track in some cases grazes the limits of this uncertainty. We expect that the agreement will improve when we do a better job of reducing the orbital eccentricity and the various subtle sources of numerical truncation error, however such calculations are still ongoing. We cannot say for certain that we have found a ``correct'' measure of black hole spin axis for all systems, however we can say that we have drastically reduced spurious features that arise from other measures. 

We note that the improved nutation behavior of a spin axis computed from $\vec \phi_i$ might imply improved behavior in analytical and surrogate models matched to numerical relativity results. We encourage further analysis along these lines. 

As a final note, it might be considered imprudent to draw broad conclusions about the behavior of spin axis measures from the one {\em specific} system studied here (5:1 mass ratio, smaller hole nonspinning, larger hole initially spinning about the axis connecting the holes), however we expect the features studied here to apply to a much broader range of cases. In particular, the ``spurious nutations'' that have been our focus --- oscillations of the spin axis along the orbital plane, which are precisely zero in the PN calculations included here --- are well-defined and one would expect them to decouple from initial spin components perpendicular to the orbital plane. A more interesting question is whether the spin measure supported by this research retains its quality of agreement with post-Newtonian expectations in cases where {\em both} black holes are spinning, a scenario that would involve spin-spin interactions. Because the spin-orbit based nutations, studied here, already lie near the limits of our numerical accuracy, this question will have to wait for future research.

\begin{acknowledgments}
We thank Serguei Ossokine, Leo Stein, and Aaron Zimmerman for very useful discussions and comments, as well as Elizabeth Garbee, who carried out related research when this project was at an earlier stage. The research presented here was supported by the a Cottrell College Science Award from the Research Corporation for Scientific Advancement (CCSA 23212). RO was also supported by a W.M.~Keck Fellowship in natural sciences and mathematics. Calculations were performed on the {\em Sciurus} cluster at Oberlin College, funded by the National Science Foundation (DBI 1427949), and on the {\em Wheeler} cluster at Caltech, funded by the Sherman Fairchild Foundation.
\end{acknowledgments}

\appendix
\section{Relationship between two definitions of the coordinate rotation generator}
\label{a:TheTwoGenerators}

Our treatment spin has frequently made reference to coordinate rotation generators. These are the rotation vectors associated with the preferred (up to linear transformations) coordinate system $x^i$ associated with the flat background geometry. These rotation vectors are constructed from coordinate {\em translation} vectors. Unfortunately, when the background geometry does not match the physical geometry, there are two natural ways to define coordinate translations. There are translation {\em vectors} $\vec \tau_{i} = \vec \partial_i$, whose components in some arbitrary coordinate system $x^a$ are given by the jacobian of the transformation:
\begin{equation}
\left(\vec \tau_{i}\right)^a = \frac{\partial x^a}{\partial x^i}.
\end{equation}
The $i$ index is a label denoting which of the three translation vectors
we're referring to, such as $\vec \tau_{1}$. 

There are also, however, translation {\em one-forms} $\bm{\kappa}^{i}$ whose 
components are given by the inverse jacobian:
\begin{equation}
\left(\bm{\kappa}^{i}\right)_a = \frac{\partial x^i}{\partial x^a}.
\end{equation}
By the standard properties of coordinate transformations, these two kinds of 
translation vector are {\em dual} to one another, meaning that their interior 
products satisfy:
\begin{equation}
\iota_{ \vec \tau_{j} } \bm{\kappa}^{i} = \kappa^{i}_a \tau^a_{j} = \delta^i_j.\label{e:TranslationDuality}
\end{equation}
However, the two sets of fields cannot be considered dual in the 
geometrical sense, 
because indices $a,b,c,...$ are raised and lowered with the physical spatial 
metric $g_{ab}$ while indices $i,j,k,...$ are raised and lowered with the 
flat background metric $\delta_{ij}$:
\begin{equation}
\kappa_{i}^a := \delta_{ij} g^{ab} \kappa^{j}_b  = \delta_{ij} g^{ab} \frac{\partial x^j}{\partial x^b}\neq \tau_{i}^a.
\end{equation}
Note that if the physical metric $g$ {\em were} the same as the reference metric $\delta$, simply represented in different coordinates, then one could say that $g^{ab} = X^a_k X^b_l \delta^{kl}$ (where $X^a_i := \partial x^a/\partial x^i$ is shorthand for the jacobian), and it would then follow that $\kappa_{i}^a$ would equal $\tau_{i}^a$. The inequality of these two sets of translation vectors is due to the inequality of the physical geometry and the flat reference geometry. 

Related to this inequality, it can also be noted that $\vec \tau_{i}$ and $\bm{\kappa}^{i}$ have different norms:
\begin{align}
\lvert \vec \tau_{i}\rvert^2 &= g_{ab} X^a_{i} X^b_{i} = g_{ii}\\
\lvert {\vec \kappa}_{i}\rvert^2 &= \delta_{ij} \delta_{ik} g^{ab} X^{j}_a X^{k}_b = \delta_{ij} \delta_{ik} g^{jk},
\end{align}
where no sum over $i$ is intended, and on the right-hand sides we have again 
used $X$ as a shorthand for the jacobian matrix or its inverse, and on the far 
right sides we have strained clarity of notation, placing $ijk$-indices on the 
physical metric to refer to the physical metric components evaluated in the 
reference coordinate basis. Again, these quantities would be equal if the 
background metric were the same as the physical metric, but because we need the 
reference metric to refer to a ``global'' vector space it must be flat, and it 
therefore cannot coincide with the physical metric.

Beyond the fact that $\vec \tau_{i}$ and $\bm{\kappa}^{i}$ differ locally, 
there are important differences in their natures as {\em fields}. In particular 
the translation one-forms $\bm{\kappa}^{i}$, being simple gradients of the 
background coordinates $x^{i}$, automatically have vanishing curl:
\begin{align}
{\bf d} \bm{\kappa}^{i} = {\bf d}^2 x^{i} &= 0,\\
\epsilon^{abc} \nabla_b \kappa^{i}_c &= 0, 
\end{align}
where $\nabla_b$ is the covariant derivative associated with the physical 
metric $g_{ab}$. The translation-generating {\em vectors} $\vec \tau_{i}$, 
on the other hand (which trivially equal $\vec \tau^{i} = \delta^{ij} \vec \tau_{j}$ due to the flat background metric), can have nonzero curl. 

This pattern of ``duality only in the non-metric sense'' carries over to our 
two definitions of coordinate {\em rotations}:
\begin{align}
\vec \phi_{i} &:= \eta_{ij}{}^k \delta x^j \vec \tau_k, \label{e:vecphidef}\\
\bm{\varphi}^{i} &:= \eta^i{}_{jk} \delta x^j \bm{\kappa}^k.\label{e:formphidef}
\end{align}
Here, $\delta x^j := x^j - x_0^j$ represents the position, in the background 
coordinate system, relative to the coordinate centroid $x_0^j$. The duality 
relationship between $\vec \tau_{i}$ and $\bm{\kappa}^{i}$ imply a 
relationship between these rotation generators, familiar from the case 
of flat space:
\begin{align}
\iota_{\vec \phi_{j} } \bm{\varphi}^{i} = \varphi^{i}_a \phi^a_{j} &= \eta^i{}_{kl} \eta_{jm}{}^n \delta x^k \delta x^m \kappa^{l}_a \tau^a_{n}\\
&= \eta^i{}_{kl} \eta_{jm}{}^n \delta x^k \delta x^m \delta^l_n\\
&= \eta^i{}_{kn} \eta_{jm}{}^n \delta x^k \delta x^m\\
&= \delta^i_j \lvert \vec {\delta x}\rvert^2 - \delta x^i \delta x_j.
\end{align}

As for the translation vectors, this simple duality relationship does not mean 
the the two rotation generators are equal in a geometrical sense. It is 
straightforward to show that $\phi_{i}^a \neq g^{ab}\delta_{ij} \varphi^{j}_b$, 
their norms differ, and their derivatives differ.

For our analysis of boost- and centroid-dependence of the spin calculated from 
these coordinate rotation generators, we need to compute their 2-surface 
divergence and curl, after projecting them down to the 2-surface. This task 
is easiest for $\bm{\varphi}^{i}$, because its geometrical index arises simply 
from a partial derivative operator. If we introduce a coordinate system $x^a$ 
where the first element is constant on the 2-surface while the other two vary, 
then the projection is accomplished by a simple change of index letter:
\begin{equation}
\varphi^{i}_A = \eta^i{}_{jk} \delta x^j \partial_A x^k.
\end{equation}
The 2-surface divergence of this is simply:
\begin{equation}
D^A \varphi^{i}_A = \eta^i{}_{jk} \left(D^A x^j\right) \left(D_A x^k\right) + \eta^i{}_{jk} \delta x^j D^2 x^k,
\end{equation}
where we have noted that the global coordinates $x^i$ are treated as scalars with respect to the surface covariant derivative. Also, when taking derivatives of $\delta x^k = x^k - x_0^k$, the centroid coordinates $x_0^k$ disappear because they are taken as constants. The first term above vanishes due to index symmetries and the remainder is:
\begin{equation}
D^A \varphi^{i}_A = \eta^i{}_{jk} \delta x^j D^2 x^k.\label{e:DivFormBasedGen}
\end{equation}

The surface curl of the $\bm{\varphi}^{i}$ can be computed similarly:
\begin{eqnarray}
\epsilon^{AB} \nabla_A \varphi^{i}_B &=& \eta^i{}_{jk} \epsilon^{AB} D_A \left( \delta x^j D_B x^k \right)\\
&=& \eta^i{}_{jk} \epsilon^{AB} \left(D_A x^j\right) \left(D_B x^k\right) \nonumber \\
& & + \eta^i{}_{jk} \delta x^j \epsilon^{AB} D_A D_B x^k.
\end{eqnarray}
Now the final term vanishes, because the surface covariant derivative has zero torsion, leaving simply:
\begin{equation}
\epsilon^{AB} \nabla_A \varphi^{i}_B = \eta^i{}_{jk} \epsilon^{AB} \left(D_A x^j\right) \left(D_B x^k\right).\label{e:CurlFormBasedGen}
\end{equation}
Note that the centroid coordinates $x_0^i$ (implicit in $\delta x^i$) do not appear in the curl, a point that implies a close relationship between the boost-gauge fixing and centroid fixing for spin based on the $\bm{\varphi}^{i}$ rotation generators, as described in Sec.~\ref{s:Translation}.

The calculation of surface derivatives of the other rotation generators, $\vec \phi_{i}$, is somewhat more tedious, requiring explicit projection onto the 2-surface, with projector:
\begin{equation}
h^a_b := \delta^a_b - s^a s_b,
\end{equation}
where $\hat s$ is the unit surface normal. The surface covariant derivative can be computed by projection of the spatial covariant derivative $\nabla_a$, associated with the physical spatial metric. The divergence is:
\begin{eqnarray}
D_A \phi^A_{i} &=& h^b_a \nabla_b \left( h^a_c \phi^c_{i}\right)\\
&=& h^b_c \nabla_b \phi^c_{i} + \left(h^b_a \nabla_b h^a_c\right) \phi^c_{i}.
\end{eqnarray}

Plugging in the right hand side of Eq.~\eqref{e:vecphidef} for $\phi^c_i$, and simplifying, one finds that the result is:
\begin{align}
D_A \phi^A_{i} &= \eta_{ij}{}^k \left(h^a_c \partial_ax^j - k s_c \delta x^j \right)\tau^c_k\nonumber\\
 & + \eta_{ij}{}^k \delta x^j h^a_c \nabla_a \tau^c_k,\label{e:DivVecGen}
\end{align}
where $k = h^a_b \nabla_a s^b$ is the trace of the 2-surface extrinsic curvature in the 3-space. The covariant derivative of the translation vector (and its dual counterpart) can be shown to be:
\begin{align}
\nabla_a \tau^b_k &= - \tau^b_i \tau^c_k \partial_a \partial_c x^i + \Gamma^b{}_{ac} \tau^c_k,\label{e:NablaTau}\\
\nabla_a \kappa^i_b &= \partial_a \partial_b x^i - \Gamma^c{}_{ab} \kappa^i_c,\label{e:NablaKappa}
\end{align}
where $\Gamma^a{}_{bc}$ are the conventional Christoffel symbols associated with the physical spatial metric. The first terms of both of these expressions happen to vanish if one uses the same coordinates to describe the physical and background geometries, as one does in numerical codes, however to avoid subtle covariance issues we do not make that assumption here.

The surface curl of the vector-based rotation generator:
\begin{equation}
\epsilon^{AB} D_A \phi_{iB} = s_c \epsilon^{cab} \nabla_a \left(h_{bd}\phi^d_i\right),
\end{equation}
can similarly be worked out as:
\begin{align}
\epsilon^{AB} D_A \phi_{iB} &= \eta_{ij}{}^k s_c \epsilon^{ca}{}_d (\partial_a x^j) \tau^d_k\nonumber\\
& + \eta_{ij}{}^k \delta x^j s_c \epsilon^{ca}{}_d \nabla_a \tau^d_k.\label{e:CurlVecGen}
\end{align}
Note that this expression, in contrast with Eq.~\eqref{e:CurlFormBasedGen}, {\em is} dependent on the choice of coordinate center $x_0^i$ because a term involving $\delta x^i$ remains. From Eq.~\eqref{e:NablaTau}, one can see that $\nabla_a \tau^d_k$ involves Christoffel symbols that {\em are not} ``antisymmetrized away'' by the Levi-Civita tensor. The nice feature of $\bm{\varphi}^i$ mentioned just below Eq.~\eqref{e:CurlFormBasedGen}, and explored in Sec.~\ref{s:Translation}, is not shared for $\vec \phi_i$.

As a final remark, we note that it is possible to derive the simpler expressions of Eqs.~\eqref{e:DivFormBasedGen},\eqref{e:CurlFormBasedGen}, for the other rotation generators, from the expressions in Eqs.~\eqref{e:DivVecGen}, \eqref{e:CurlVecGen}, using the substitution $\tau^a_i \mapsto \delta_{ij} g^{ab} \kappa^j_b$.


\end{document}